\documentclass[useAMS,usenatbib]{mn2e}

\usepackage{graphicx}
\usepackage{amsmath}
\usepackage{booktabs}
\usepackage{journals}
\usepackage{ulem}
\usepackage{lastpage}

\title[Effect of asphericity in caustic mass estimates]
        {Effect of asphericity in caustic mass estimates of galaxy clusters}
\author[J. Svensmark et al.]
  {Jacob~Svensmark,$^1$
  Radoslaw~Wojtak,$^{1,2}$ and
  Steen H.~Hansen$^1$\\
  $^1$Dark Cosmology Centre, Juliane Mariesvej 30, 2300 \O sterbro\\
  $^2$Kavli Institute for Particle Astrophysics and Cosmology, Stanford University,\\
  \hspace{1mm}   SLAC National Accelerator Laboratory, Menlo Park, CA 94025}
\begin{document}	


\maketitle

\label{firstpage}

\begin{abstract}
The caustic technique for measuring mass profiles of galaxy clusters relies on the assumption of spherical symmetry. When applied to aspherical galaxy clusters, the method yields mass estimates affected by the cluster orientation. Here we employ mock redshift catalogues generated from cosmological simulations to study the effect of clusters intrinsic shape and surrounding filamentary structures on the caustic mass estimates. To this end, we develop a new method for removing perturbations from large-scale structures, modelled as the two-halo term, in a caustic analysis of stacked cluster data.
 
We find that the cluster masses inferred from kinematical data of $~10^{14}M_{\sun}$ clusters observed along the major axis are larger than masses from those observed along the minor axis by a factor of $1.7$ within the virial radius, increasing to $1.8$ within three virial radii. This discrepancy increases by $20\%$ for the most massive clusters. In addition a smaller but still significant mass discrepancy arises when filamentary structures are present near a galaxy cluster.

We find that the mean cluster mass from random sightlines is unbiased at all radii and their scatter ranges from $0.14$ to $0.17$ within one and three virial radii, with a $40\%$ increase for the most massive clusters. We provide tables which estimate the caustic mass bias given observational constraints on the cluster orientation.
\end{abstract}

\begin{keywords}
galaxies: clusters: general - 
cosmology: large-scale structure of Universe - 
galaxies: kinematics and dynamics
\end{keywords}

\section{Introduction} \label{sec:intro}
	Measuring masses of galaxy clusters is a problem of growing importance and difficulty at the same time. 
Accurate mass determination is necessary to make galaxy clusters a robust tool for testing cosmological models
and general relativity or constraining cosmological parameters \citep{Allen2011}. The most competitive methods to date 
comprise such techniques as cluster counts \citep{Vikhlinin2009,Rozo2010} and measurements of the growth 
rate \citep{Mantz2010,Rapetti2010} or the gas mass fraction \citep{Allen2002,Mantz2014}, although it is also worth 
mentioning a number of methods exploring secondary predictions of cosmological models such as the mass-concentration 
relation \citep{Ettori2010}, the ultimate halo mass \citep{Rines2013} and the merger contribution to cluster-sized halo growth \citep{Lemze2013}. Precise constraints on cosmological parameters are probably the main  motivation for searching for the 
most accurate mass estimators for galaxy clusters. As an example of how crucial the problem is, we recall that the recently 
reported tension in cosmological constraints determined from SZ-selected cluster counts and the Cosmic Microwave Cosmic 
temperature anisotropies based on observations of the Planck satellite may be an effect of inaccurate mass calibration of galaxy 
clusters used in the measurement \citep{Planck2013a,Planck2013b,Linden2014}. Precise constrains on dynamical and lensing mass profiles are also essential 
for testing alternative theories of gravity on cluster scales \citep{Schmidt2010,Lombriser2012}. This is probably even more 
challenging than using clusters as cosmological probes, since the predicted signals for the most likely modifications 
of standard gravity are comparable to or smaller than the typical accuracy of the current cluster mass measurements.

Most methods of cluster mass determination rely on the assumption of spherical symmetry. This seems to be consistent with the 
most commonly used operational definition of mass based on spherical overdensity which underlies most of the predictions 
from cosmological simulations. On the other hand, such an assumption is obviously in conflict with the actual shape of matter 
distribution in galaxy clusters. In contrast to cosmological simulations where spherical overdensity mass is an unambiguous 
mass definition, spherical symmetry assumed for a deprojection of an observed signal from aspherical clusters makes the mass 
measurement dependent on the actual orientation of galaxy clusters with respect to the line of sight. This is one of the major 
effects giving rise to a systematic difference between measured and actual cluster masses.

There is a growing body of observational evidence for strongly elongated shapes of galaxy clusters, both in the visual and 
dark matter component \citep{Limousin2013}. For many individual galaxy clusters, spherical symmetry has been shown 
to define an insufficiently accurate framework for an analysis of observational data and the mass inference \citep[see e.g.][]{Corless2009,Sereno2013}.
These results are not surprising, since it is well-known that simulated galaxy clusters and cluster-like 
dark matter halos are highly aspherical, with shapes approximated by preferentially prolate ellipsoids with typical major-to-minor axial ratios of $\sim 0.5$ and major-to-medium ratios of $\sim 0.7$ \citep{Allgood2006,Lemze2012}. Lack of spherical symmetry is also present in the velocity distributions which tend to be spatially 
anisotropic, with the direction of the maximum velocity dispersion aligned with the semi-major axis of the halo shape \citep{Kasun2005}, 
and axially symmetric at small radii \citep{Wojtak2013b}. This feature has been confirmed observationally using optically-selected 
clusters and massive groups of galaxies from the Sloan Digital Sky Survey \citep{Ski2012,Wojtak2013a}. The recovered axial symmetry of the velocity 
distributions of galaxies in clusters turned out to be essential in reconciling mass estimates from abundance matching and from the velocity 
dispersions.

Despite a number of attempts of employing aspherical models for the matter distribution in the analysis of X-ray and lensing 
observations of galaxy clusters \citep[see e.g.][]{Corless2009,Samsing2012,Sereno2013}, all dynamical methods are based on spherical symmetry. 
As the first step in addressing the problem of asphericity in mass measurements based on kinematics of galaxies in clusters, we study how 
dynamical mass estimators assuming spherical symmetry depend on the orientation of galaxy clusters with respect to the line of sight.  
We assess this effect by studying dynamical masses inferred from mock kinematic data of galaxy clusters generated from cosmological 
simulations. We restrict our analysis to dynamical masses measured with the so-called caustic technique \citep{Diaferio1999} which 
is one of the commonly used methods of mass determination in galaxy clusters \citep[see e.g.][]{Biviano2003,Rines2006,Lemze2009,Geller2013,Rines2013}. 
The caustic technique does not explicitly assume dynamical equilibrium beyond the virial radius, therefore it can be used to measure masses of galaxy clusters at distances larger than their virial radius and allows us to study the mass bias in a wide range of radii. 
As all dynamical methods currently applied to cluster data, it assumes spherical symmetry and testing this 
assumption is an objective of this work. Dependence of the mass measurement on the orientation of galaxy clusters with respect to the 
sight line is expected not only due to clusters' intrinsic phase-space shapes, but also due to co-alignment of the surrounding large-scale 
structures. The latter effect has been clearly shown both in cosmological simulations \citep{Libeskind2013} and observations 
\citep{Paz2011}, and it is expected to modulate the contribution of background galaxies in kinematic samples and thus to 
affect the final estimate of dynamical mass.

The paper is organized as follows. Section \ref{sec:data} describes the treatment of data from an $N$-body dark matter particle simulation, and of selected halos within it. Section \ref{sec:stacks} describes the rotation and stacking of the clusters in ellipsoidal, filamentary and spherical stacks. Section \ref{sec:caustic_technique} outlines the theory of the caustic technique for mass estimation of galaxy clusters. Section \ref{sec:results} presents the caustic mass estimates of the stacks as obtained by using different lines of sight, and Section \ref{sec:discussion} discusses how this works as a bias in mass estimation. Section \ref{sec:conclusion} sums up the conclusions of the work done in this paper.
\section{Simulations and mock catalogues} \label{sec:data}
	The caustic technique of mass estimation takes as input the projected phase-space data of an observed galaxy cluster, i.e. the projected sky positions and line of sight velocities of its member galaxies. In order to apply and evaluate the performance of the caustic technique in spatially isotropic and anisotropic settings we set out to compile a range of mock phase-space diagrams of simulated galaxy clusters.  We assume that realistic representations of phase-space diagrams of clusters can be obtained by considering just dark matter particles from $N$-body simulations. The particle data were obtained from the Bolshoi simulation\footnote{The simulation is publicly available through the Multi Dark database (http://www.multidark.org). See \cite{Riebe2013} for details of the database.}, which simulates $2048^3$ dark matter particles, each with a mass of 1.35$\times 10^{8}h^{-1}\,M_{\sun}$ \citep{klypin2011}. The simulation evolves from redshift $z=80$ inside a box volume of side length $250\,h^{-1}\mathrm{Mpc}$ and uses cosmological parameters consistent with measurements based on the WMAP five-year data release \citep{komatsu2009} and the abundance of optical clusters from the Sloan Digital Sky Survey \citep{Rozo2010}, $\Omega _m = 0.27$, $\Omega _\Lambda = 0.73$, $\sigma_8=0.82$ and $h=0.7$ such that $H_0=100\,h\,$km$\,$s$^{-1}$Mpc$^{-1}=70\,$km$\,$s$^{-1}$Mpc$^{-1}$.

We construct the phase-space diagrams by using two sets of data from the Bolshoi simulation at redshift $z=0$ in conjunction, namely a location and velocity subset of $8.6\times 10^6$ randomly drawn dark matter particles, along with a Bound Density Maximum (BDM) halo catalogue also obtained from the Multidark database, which lets halos extend up to an overdensity limit of $360\,\rho _{\rm b}$, where $\rho_{\rm b}$ is the background density. Throughout, we shall refer to the former dataset when mentioning particles, and the latter when mentioning halos. For each halo in the catalogue all particles were assigned an additional radial Hubble flow velocity $v_h = 100\,h\,r\,$km$\,$s$^{-1}$Mpc$^{-1}$ according to their 3-dimensional distance $r$ to the halo center. We choose to use the exact halo centers of the BDM catalogue here and throughout our analysis in order to provide optimal conditions for the application of caustic technique. To ensure comparable proportions of the halos, the positions of their member particles were normalized by the virial radius $R_v$ and their velocities by the virial velocity $V_{v}=\sqrt{G\,M_{v}/R_{v}}$, where $M_{v}$ is the virial mass. All of these quantities along with the halo center locations are provided in the BDM halo catalogue. Clusters were chosen within two mass bins, namely in the range of $M_v \in \left[ 1,2 \right]\times 10^{14}h^{-1}M_{\sun}$, which yielded 230 distinct halos in the catalogue, and $M_v > 2\times 10^{14}\,h^{-1}M_{\sun}$ which yielded 101 distinct halos in the catalogue. The two shall throughout be referred to as the 'low mass bin' and the 'high mass bin' respectively.

\section{Stacking the data}\label{sec:stacks}
Two big sources of spatial anisotropy in clusters stems from elongation of the cluster itself (intrinsic shape) and surrounding large scale structure i.e. filaments, walls and voids. In order to isolate these morphological features for further analysis the halos from each mass bin were arranged concentrically in three separate stacks. For each stack all halos and their particles were rotated individually to align and isolate the geometric features of elongation and filamentary structure. This yielded for each mass bin:
\begin{itemize}
  \item An \textit{ellipsoidal stack} where halos were modelled as ellipsoidal structures and rotated so that the three principal axes of each halo were aligned
  \item A \textit{filamentary stack} that aligned the direction of largest filament associated with each halo
  \item A \textit{spherical stack} for reference with arbitrary orientation of each halo
\end{itemize}
The three configurations made it possible to choose any line of sight through the anisotropic stacks and compare mass estimates from caustics with those of the \textit{spherical stack}. Because the stacks differ only in orientation of individual halos, they have the same true cumulative mass profile $M(<r)$. Therefore any difference in caustic mass estimation between the \textit{ellipsoidal} or \textit{filamentary stack} and the \textit{spherical stack} expresses an anisotropy bias in the caustic method of mass estimation. Considering stacks rather than individual clusters separately allows for a clear disentangling of the shape-orientation effect from such effects as amount of substructures or degree of dynamical equailibrium. The latter effects may affect the mass determination for individual galaxy clusters, but they are averaged out for stacks.

\subsection{Ellipsoidal geometry and alignment}
To create a smooth halo with clearly pronounced triaxiality and little interference from cluster substructure, ongoing merging or large scale structure, each of the halos were rotated according to their directions of elongation and placed in a stack. As a measure of elongation and orientation we invoked an ellipsoidal model by considering the shape tensor of each halo. Its three eigenvalues $\lambda_i$ define the principal axes of the ellipsoid, and its eigenvectors define their orientation. For each halo the shape tensor $S_{jk}$ \cite[see e.g.][]{Zemp2011} can be obtained by summing over its $N$ member particles within some radius:
\begin{equation}
	S_{jk} = \sum_{i=1}^{N}(x_j)_i(x_k)_i.
\end{equation}
Here $x_j$ and $x_k$ are the $j$'th and $k$'th components of the 3-dimensional position vector of the $i$'th particle. We use particles within $R_{v}$ to avoid interference from any nearby large-scale structure. The eigenvalues $\lambda_a$, $\lambda_b$ and $\lambda_c$ of $S_{jk}$  then give the principal axes of the ellipsoid:
$a=(\lambda _a)^{1/2}$, 
$b=(\lambda _b)^{1/2}$ and 
$c=(\lambda _c)^{1/2}$. 
The eigenvalues are ranked so that $a\geq b \geq c$. Their corresponding eigenvectors dictate the direction of each principal axis. The three semi-axes $a$, $b$ and $c$ of each cluster were aligned with the $x$-, $y$- and $z$-axis respectively. This configuration is sketched in the top part of Fig. \ref{fig:model}, and it shall be referred to as an \textit{ellipsoidal stack}.

For each of the halos, the triaxiality parameter $T$ \citep{franx1991} is defined as 
\begin{equation}
T=\frac{a^2-b^2}{a^2-c^2}.
\end{equation}
Clusters with $T=0$ are purely oblate, those with $T=1$ are purely prolate. The triaxiality parameter will assist in dividing the clusters from each mass bin into ellipsoidal substacks binned according to $T$. This allows for an examination of the properties of predominantly oblate or prolate clusters. After stacking all halos in each mass bin the shape tensor was calculated within $R_{v}$ for the stacks as a whole, yielding for $b/a=0.78$, $c/a=0.65$ and $T=0.69$ for the low mass bin, and $b/a=0.76$, $c/a=0.63$ and $T=0.70$ for the high mass bin. Consequently this is a significantly aspherical prolate configuration for both mass bins. These numbers are also summarized in the 'Triaxiality' columns of Table \ref{tab:results}.

\subsection{Filament geometry and alignment}\label{sec:filament_geo}
\begin{figure}
	\begin{center}
		\includegraphics[width=8.3cm]{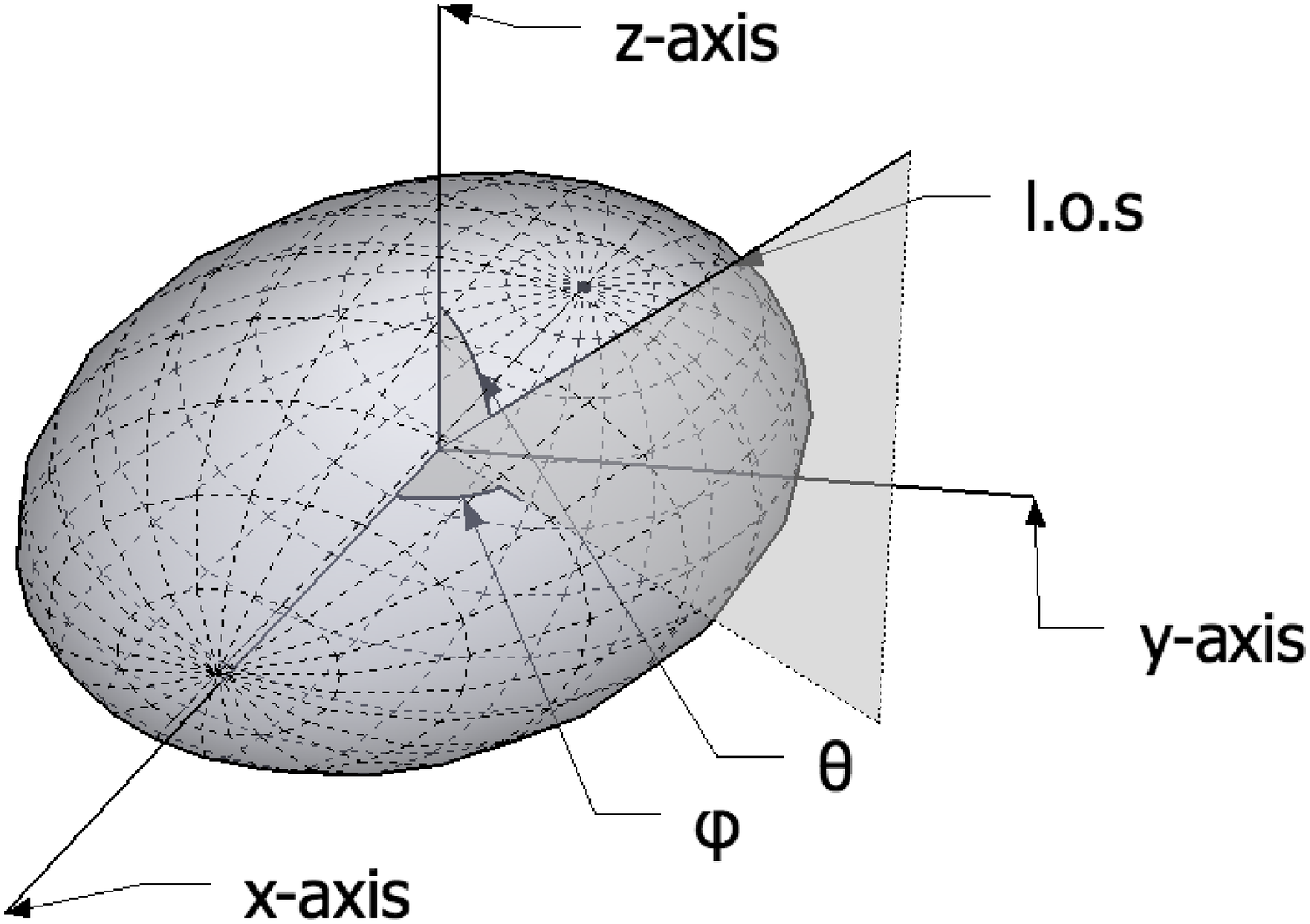}
		\includegraphics[width=8.3cm]{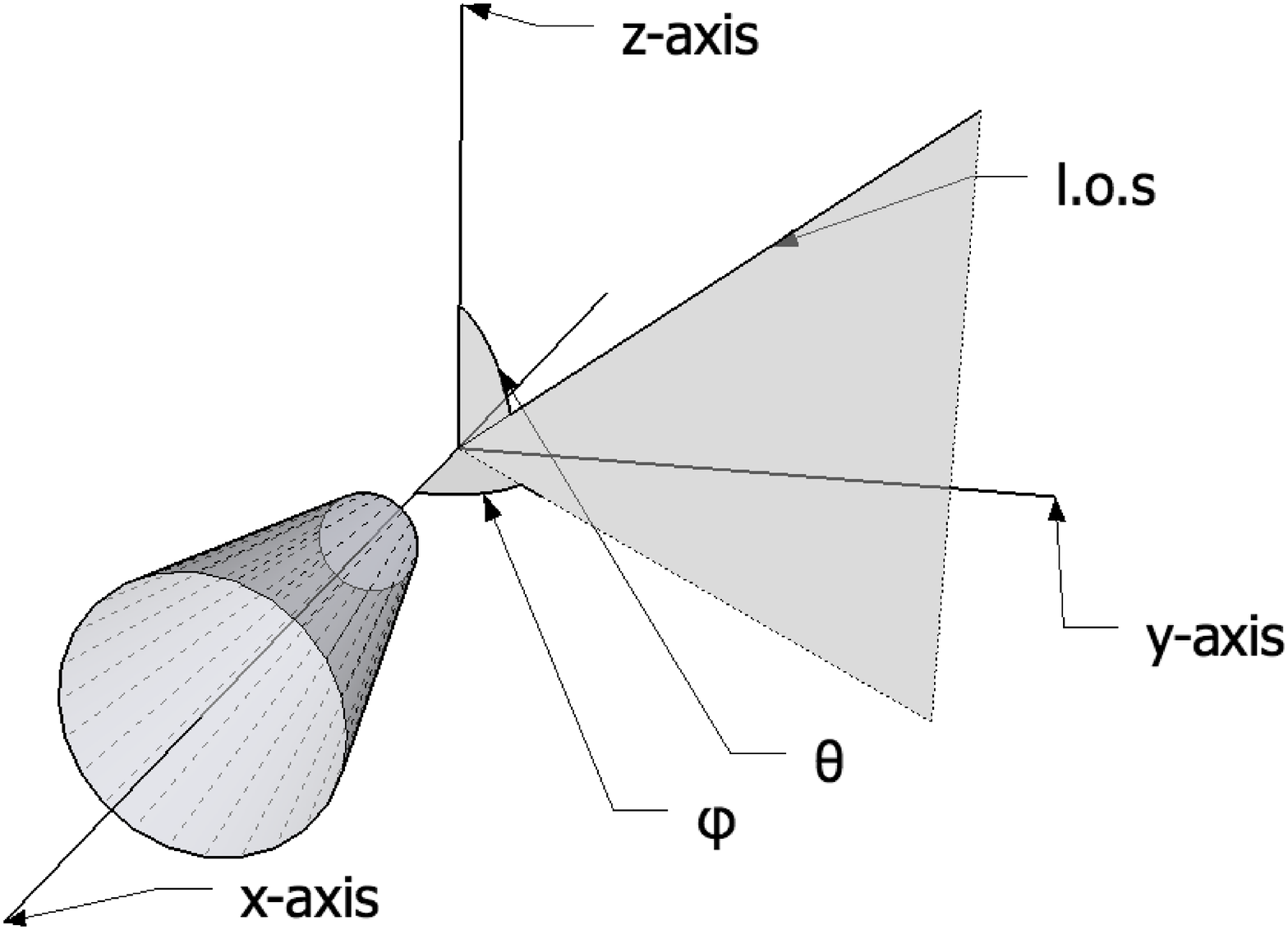}
		\caption{\textit{Top:} Sketch showing the geometry of an \textit{ellipsoidal stack}. Note how the ellipsoid has its longest major axis $a$ along the $x$-axis, the shorter $b$ along $y$ and the shortest $c$ along the $z$-axis. \textit{Bottom:} Sketch showing the geometry of a filamentary stack. Notice the placement of the 30$^{\circ}$ cone frustum (filament) along the positive $x$-axis. The 'l.o.s.' arrow indicates the line of sight. Note that both figures are not to scale.}
		\label{fig:model}
	\end{center}
\end{figure}

Filamentary structure may manifests itself as not necessarily straight strings of material between galaxy clusters. As a simple approach to create a smooth halo with a pronounced filament associated with it, we rotate each individual halo such that the largest filament associated with each halo (if any) is aligned with the positive $x$-axis. As a measure of filamentary structure the direction of highest particle number density as seen from the center of the halo was used. Particles between $1\,R_{v}$ and $5\,R_{v}$ were examined for each halo. The direction of maximal number density within a cone frustum of opening angle 30$^{\circ}$ was located. The particles from each halo were again placed concentrically in a stack, and rotated such that the maximal density direction was aligned with the positive $x$-axis. In this way the stack puts emphasis on surrounding filament structure. Because of the axisymmetric geometry of the cone frustum there are no preferred directions within the plane orthogonal to the filament, and consequently two random (orthogonal) directions were chosen within this plane for the $y$- and $z$-axis for each halo. This configuration is sketched in the bottom part of Fig. \ref{fig:model}, and it shall be referred to as a \textit{filamentary stack}. Note that only the filament frustum and not the core of the cluster is sketched in this figure, even though naturally still present. Because the stack is oriented after surrounding filaments, we expect the inner parts of it to be fairly spherical, although some alignment of ellipsoidal clusters principal axis and filaments have been reported \citep{hahn2007,Libeskind2013}. Using a cone frustum as a geometric model of filaments in a stack will produce a very straight 'filament', which is true for some but not necessarily all individual clusters \citep{colberg2005}.
After stacking all halos according to their filamentary structure, the shape tensor for stack particles within $R_{v}$ had $b/a=0.89$, $c/a=0.86$ and $T=0.81$ for the low mass bin, and $b/a=0.88$, $c/a=0.85$ and $T=0.81$ for the high mass bin. This configuration is much more spherical (within $R_{v}$) than that of the \textit{ellipsoidal stack}, though some triaxiality is still present.

\subsection{Spherical reference geometry}\label{sec:sphere_stack}
When testing for effects of anisotropy it is good to have an ideal spherical configuration as a reference point. To create a spherically symmetric halo, the individual halos in each mass bin were all superposed 10 times each in a third stack, with a new random rotation for each superposition. This yielded a highly spherical configuration, where triaxiality of clusters and filamentary structure is smoothed out on average. It shall be referred to as a \textit{spherical stack}. With these three stack types in hand it is possible to investigate ellipsoidal cluster properties along different lines of sight using the spherical cluster properties as reference. The shape tensor for particles within $R_{v}$ for this stack yielded $b/a=1.00$ and $c/a=0.99$ in the low mass bin and $b/a=0.98$ and $c/a=0.97$ for the high mass bin. Both are thus very close to a spherical distribution, in which case $T$ is undefined.

\section{The caustic technique}\label{sec:caustic_technique}
	The escape velocity of a spherical gravitationally bound system of particles relates to its gravitational potential through
\begin{equation}
\label{eq:escapevelocity}
	v^2_{esc}(r) = -2\Phi (r),
\end{equation}
where $r$ indicates the 3-dimensional distance to the center of the system. Since the potential is a non-decreasing function of $r$, one would expect to find fast moving objects closer to the cluster center, and gradually slower objects further out, with the maximally observed velocity defined by $v_{esc}(r)$.

Following the work of \cite{Diaferio1999} we conduct an analysis of our stacked cluster particle data using the now standard caustic technique for mass estimation of galaxy clusters. The idea of the caustic procedure is to locate the amplitude of a caustic envelope which encloses bound particles placement in the projected ($R$-$v$) phase-space, and relate it to the escape velocity. Here, the observables $R$ and $v$ indicate the projected distance to the cluster center and the line of sight velocity respectively. The escape velocity profile relates directly to the potential which may then be integrated to find the contained mass profile of the cluster. To do so, one must assume sphericity. Because of our stacking approach, an additional correction for surrounding large-scale structures must be invoked. Note that the following sections rely heavily on derivations presented in \cite{Diaferio1999}, \cite{serra2011} and \cite{gifford2013_1}.

\subsection{Phase space density distribution}\label{sec:kernel}
Different authors take slightly different approaches to determining the actual caustic amplitude from a set {($R$,$v$)} of observed particles in projection. The usual approach involves estimating the underlying density distribution $f(R,v)$ by using each particle as a tracer in a kernel density distribution estimation. Let $(R,v)$ describe any position in the projected ($R$-$v$) phase space, and let $(R_i,v_i)$ describe the location of the $i$'th particle tracer. We estimate the local density distribution at $(R,v)$ by summing over all $N$ particle tracers in our stack:
\begin{equation}\label{eq:dist}
f(R,v) = \frac{1}{N}\sum_{i=1}^{N}\frac{1}{h_Rh_v}K\left(\frac{R-R_i}{h_R},\frac{v-v_i}{h_v}\right).
\end{equation}
Here, $h_R$ and $h_v$ control the width of the kernel smoothing in the $R$ and $v$ directions respectively. Different forms of the kernel $K$ will naturally yield different $f(R,v)$. \cite{Diaferio1999} uses an adaptive kernel, whereas \cite{gifford2013_1} use a Gaussian type kernel, which they argue introduces no significant error to the distribution estimate. We shall adapt the latter Gaussian kernel, which takes the form:
\begin{equation}
K(x_R,x_v) = \frac{1}{2\pi h_R h_v}\exp\left[-\frac{\left( x_R^2+x_v^2 \right)}{2}\right].
\end{equation}
We set $h_R = N^{-\frac{1}{6}}\sigma_R$ and $h_v = N^{-\frac{1}{6}}\sigma_v$ as the rule-of-thumb optimal size of the kernel, where $\sigma _R$ and $\sigma _v$ are the dispersions in the $R$- and $v$-direction respectively. For more information on these choices see statistical work by \cite{silverman1986}, and cosmological work by \cite{pisani1993}. 
The top panel of Fig.~\ref{fig:dens_dist} shows an example of the phase-space density calculated for the spherical stack.

\subsection{2-halo term correction}
The key idea of the caustic technique is to associate contours of constant $f(R,v)=\kappa$ with potential caustic amplitudes $\mathcal{A}_{\kappa}(R)$. 
In the general case of individual galaxy clusters, iso-phase-space-density contours at large projected distances are not symmetric with respect to the cluster 
bulk velocity and often enhanced by nearby large-scale structures. A common way to circumvent this projection effect is to choose the minimum of the iso-density contour determined for positive and negative velocities (upper and bottom half of the phase-space diagram) for a given $\kappa$. This procedure minimizes the effect of perturbation 
due to the presence of nearby large-scale structures, however it does not eliminate it completely. Projection effects become difficult to deal with when perturbing large-scale structures are distributed nearly symmetrically with respect to the cluster bulk velocity. In this case, the observationally determined caustic curves 
often exhibit unphysically flat profiles extending up to $8$ h$^{-1}$Mpc \cite[see][]{Rines2003,Rines2006}. On the other hand, if perturbing structures are located 
asymmetrically (more likely to happen) then the minimum amplitude method yields accurate caustic lines.

Phase-space stacking makes the contribution from nearby large-scale structures symmetric in velocity space. This excludes a possibility of using the minimum 
amplitude approach as a method for removing perturbations from nearby large-scale structures. However, the contribution from nearby structures 
can be taken into account in a statistical way by considering a weighting function describing the probability of selecting halo particles given the projected 
radius and velocity. The weighting function depends in general on radius and velocity. However, in the velocity range dominated by peculiar velocities it 
can be approximated by
\begin{equation}
w(R)=\frac{\rho_{1h}(R)}{\rho_{1h}(R)+\rho_{2h}(R)},
\label{weighting_function}
\end{equation}
where $\rho_{1R}$ is the halo density profile and $\rho_{2h}$ is the so-called two-halo term describing the contribution from overlapping large-scale 
structures. The phase-space density corrected for the contribution from overlapping large-scale structures is then given by
\begin{equation}
\hat{f}(R,v)=w(R)f(R,v).
\label{weighted_f}
\end{equation}
The caustic amplitude is determined using the corrected $\hat{f}(R,v)$.

We calculated the weighting function approximating $\rho_{1h}$ by an NFW profile with the virial mass and the concentration parameter equal 
to their mean values calculated in each cluster bin. The two-halo term was computed using the following formula \citep{Hayashi2008}
\begin{equation}
\rho_{2h}(r)=\rho_{b}b^{L}_{m}(M)\xi^{L}_{m}(r),
\end{equation}
where $\rho_{b}$ is the background density, $b^{L}_{m}$ is the linear halo bias and $\xi^{L}_{m}(r)$ is the linear correlation function given by
\begin{equation}
\xi^{L}_{m}=\frac{1}{2\pi^{2}}\int \textrm{d}k k^{2}P(k)\frac{\sin(kr)}{kr},
\end{equation}
where $P(k)$ is the linear power spectrum. The halo bias was estimated using approximations provided by \citet{Tinker2010} and the power 
spectrum was computed for cosmological parameters used in the Bolshoi simulation.

The bottom panel of Fig.~\ref{fig:dens_dist} shows the phase-space density $\hat{f}(R,v)$ for the spherical stack, after correction using the two-halo term. 
It is clearly visible that the weighting function reverses a divergent trend of iso-density of the uncorrected phase-space density at large radii 
(top panel) which occurs due to the presence of overlapping large-scale structures. As we shall demonstrate in the following section, the 
caustic amplitude profile determined from the corrected phase-space density $\hat{f}(R,v)$ accurately recovers the true mass profile. Neglecting this 
correction does not have an effect on the caustic and the mass determination at small radii $r<R_v$, but it leads to a noticeable mass overestimation 
at large radii: by 20 per cent at $2\,R_v$ and by 50 per cent at $3\,R_v$.

The outlined procedure for correcting the observed phase-space density $f(R,v)$ is an analogue to the minimum amplitude approach applied 
to phase-space diagrams of individual galaxy clusters. In both cases, the goal is to remove the contribution from overlapping 
large-scale structures. Our work is based on stacked data and we therefore employ the same weighting procedure given by eq.~(\ref{weighted_f}) 
in all considered cases. We use two weighting functions calculated separately for two cluster mass bins.

\subsection{Determination of the caustic profile}

The caustic amplitude at projected radius $R$ is related to the escape velocity of the cluster by
\begin{equation}
\left<v^2_{esc}\right>_{R,\kappa} = \int_0^R\mathcal{A}_{\kappa}^2(r)\phi (r)dr/\int_0^R\phi (r)dr,
\end{equation}
where $\phi (r) =\int \hat{f}_{\kappa}(r,v)dv$. The caustic amplitude $\mathcal{A}(r)$ is then chosen as the $\hat{f}(r,v)=\kappa$ that minimizes 
\begin{equation}\label{eq:calibration}
M(\kappa ,R) = \left| \left<v^2_{esc}\right>_{R,\kappa} - 4\left<v^2\right> _{R}\right| ^2
\end{equation}
within the virial radius by setting $R=R_{v}$. For more information on this minimization see \cite{gifford2013_2}. For $\left<v^2\right>_{R}$ we use the mass-weighted mean line of sight velocity dispersion inside $R=R_{v}$. When observing naturally $R_{v}$ is not know a priori, and therefore an iterative scheme must be applied. An initial guess of $R_v$ results in a caustic mass $M_v$, which converts to a new virial radius $\propto M_v^{1/3}$, with which the caustic technique can be re-applied iteratively until convergence on the final caustic amplitude and observed virial radius $R_{v,obs}$. We apply this scheme in the appendix with an initial guess of $1\,R_v$, however throughout most of this work we assume the true $R_v$ to be known. Finally, we limit the caustic amplitude such that if $\mathrm{d}\ln{\mathcal{A}/\mathrm{d}\ln{r}}>\zeta$ we impose a new value for $\mathcal{A}$ such that $\mathrm{d}\ln{\mathcal{A}/\mathrm{d}\ln{r}}=\zeta$. Following \cite{serra2011} we choose $\zeta = 2$, to only cap the very rapid and non-physical increases in the escape velocity, although some authors chose lower values \citep{Diaferio1999,Lemze2013}. This yields a final caustic amplitude $\mathcal{A}(r)$, which will give the caustic mass profile in the next section.

\subsection{Caustic Amplitude and Gravitational Potential}
With a measure of the caustic amplitude in hand, the usual approach is to relate it to the potential profile of the system. \cite{Diaferio1999}, \cite{serra2011} and \cite{gifford2013_1} implement some form of the equation
\begin{equation}\label{eq:causticequation}
-2\Phi(r)=g(\beta)\mathcal{A}^2(r),
\end{equation}
where $g(\beta)=\frac{3-2\beta}{1-\beta}$ and $\beta(r) = 1 - \left<v_{\theta}^2+v_{\phi}^2\right>  /2\left<v_{r}^2\right>$ is the velocity anisotropy parameter. Here $v_{\theta}$, $v_{\phi}$ and $v_r$ are the longitudinal, azimuthal and radial components of the 3-dimensional velocity, and brackets indicate the average over velocities in the volume d$^3r$ at position $r$. We write the infinitesimal mass element for a sphere of density $\rho(r)$ in the form
\begin{equation}
G\,\mathrm{d}m = -2\Phi (r) \mathcal{F}(r)\mathrm{d}r,
\end{equation}
where $\mathcal{F}(r)=-2\pi G\rho (r)r^2/\Phi (r)$. Using equation (\ref{eq:causticequation}) in the above and integrating we get the cumulative mass profile as a function on $\mathcal{A}(r)$, $\mathcal{F}(r)$ and $g(r)$:
\begin{equation}
GM(<R) = \int _0 ^R \mathcal{F}(r)g(r) \mathcal{A}^2(r)\mathrm{d}r.
\end{equation}
\cite{Diaferio1999} argues that the product $\mathcal{F}_{\beta}=\mathcal{F}(r)g(r)$ varies slowly with $r$, and can be taken as constant, which finally relates the caustic amplitude directly to the cumulative mass profile as
\begin{equation}\label{eq:massprofile}
GM(<R) =\mathcal{F_\beta} \int _0 ^R \mathcal{A}^2(r)\mathrm{d}r.
\end{equation}
The choice of $\mathcal{F}_{\beta}$ varies for different authors. We use a value of $\mathcal{F}_{\beta}=0.59$ for the low mass bin, and $\mathcal{F}_{\beta}=0.63$ for the high mass bin, which is in line with previous work \citep[e.g.][]{Diaferio1999,serra2011,gifford2013_1}, which is slightly low compared to  Fig. 7 of \citep{Biviano2003}, however in good agreement with Fig. 11 of \citep{Biviano2013} and Fig. 9 of \citep{Lemze2009}. Using these values we recover the true cumulative mass profiles using the \textit{spherical stacks} excellently up to within $3\,R_{v}$, as shall be discussed in later sections of this paper (see bottom of Fig. \ref{fig:caustic_amplitudes}).

\begin{figure}
	\begin{center}
\hbox{\hspace{0.4cm}\includegraphics[width=8.8cm]{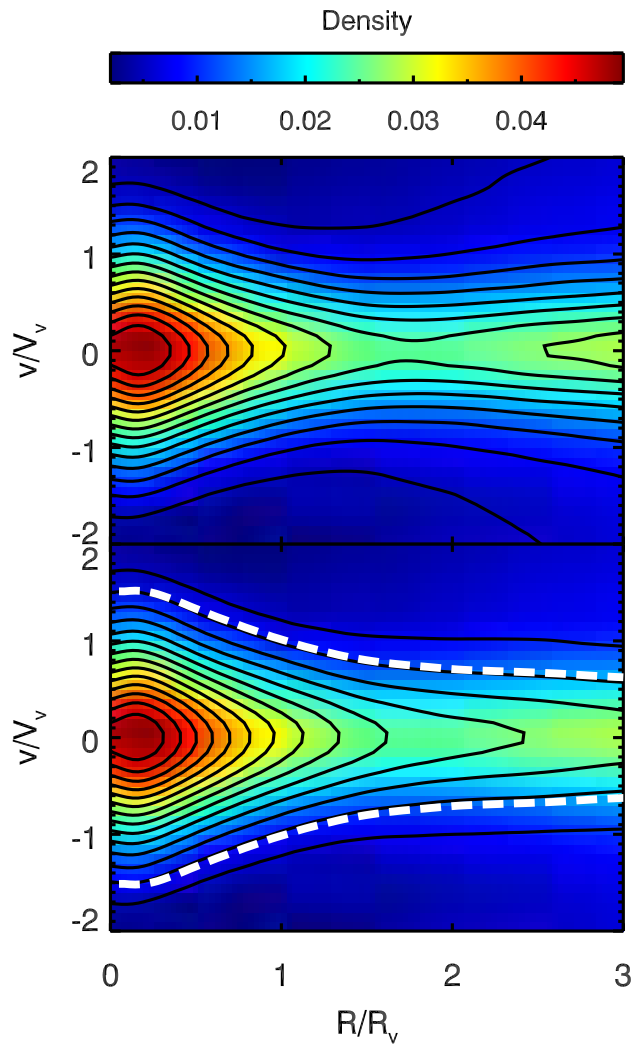}}
		\vspace{-1cm}
		\caption{Top: Projected phase-space density $f(R,v)$ of DM particles of the 230 clusters from the low mass bin \textit{spherical stack} as obtained by the Gaussian kernel. Bottom: Projected phase-space density show in the top panel after correction for the two-halo term, $\hat{f}(R,v)$. Black lines indicate isodensity contours, and the white dashed lines indicates caustic amplitude selected by the caustic technique.}	
		\label{fig:dens_dist}
	\end{center}
\end{figure}

In short the distribution of particles in projected phase space as obtained by equation (\ref{eq:dist}), and the velocity dispersion entering in equation (\ref{eq:calibration}) work together to define and select the caustic amplitude. It is precisely the interplay between these two mechanisms that spatial anisotropy is expected to affect, such that the inferred mass profile from aspherical clusters obtained with equation (\ref{eq:massprofile}) might deviate from the spherical case.
\section{Results}\label{sec:results}
	To quantify how the geometry of clusters affects the amplitude selected by the caustic technique for different projections, both the \textit{spherical}, \textit{ellipsoidal} and \textit{filamentary stack} were projected using varying lines of sight for both mass bins under consideration. The bottom part of Fig. \ref{fig:dens_dist} shows the density distribution $\hat{f}(R,v)$ as calculated by the kernel density estimation technique described in Section \ref{sec:kernel} and corrected by the 2-halo term. The distribution is calculated from the low mass bin \textit{spherical stack} for an example choice of line of sight at $(\theta,\phi)=(90^{\circ},0^{\circ})$. The black isodensity contours show the classical trumpet shape along which we expect the caustic amplitude to lie. The white dashed lines on the plot shows the actual caustic amplitude for this distribution as found by the caustic method described above.

In the top panel of Fig. \ref{fig:caustic_amplitudes} the curves which are symmetric around $v=0$ show caustic amplitudes found for the low mass bin \textit{ellipsoidal stack} for 4 lines of sight with directions as indicated on the bottom panel. The outer red long dashed curves show the caustic amplitude found using a sight line along the semi-major axis of the stack, i.e. $(\theta,\phi)=(90^{\circ},0^{\circ})$. The inner blue dash-dot-dot-dotted curves show the caustic amplitude found from a sight line along the semi-minor axis of the stack, ie. $(\theta,\phi)=(0^{\circ},0^{\circ})$. The two curves in between show caustic amplitudes using sight lines $\theta =60^{\circ}$ (green short dashed) and $\theta =30^{\circ}$ (cyan dash-dot) (see legend in the bottom plot of Fig. \ref{fig:caustic_amplitudes}). The bottom panel of Fig. \ref{fig:caustic_amplitudes} shows the cumulative mass profiles $M(<R)$ as obtained from the caustic amplitudes shown in the top panel combined with equation (\ref{eq:massprofile}). The red long dashed, green short dashed, cyan dash-dotted and blue dash-dot-dot-dotted lines are thus from the low mass bin \textit{ellipsoidal stack} with line of sight as indicated in the legend, keeping $\phi=0^{\circ}$.

\begin{figure*}
	\begin{center}
		\includegraphics[width=17	cm]{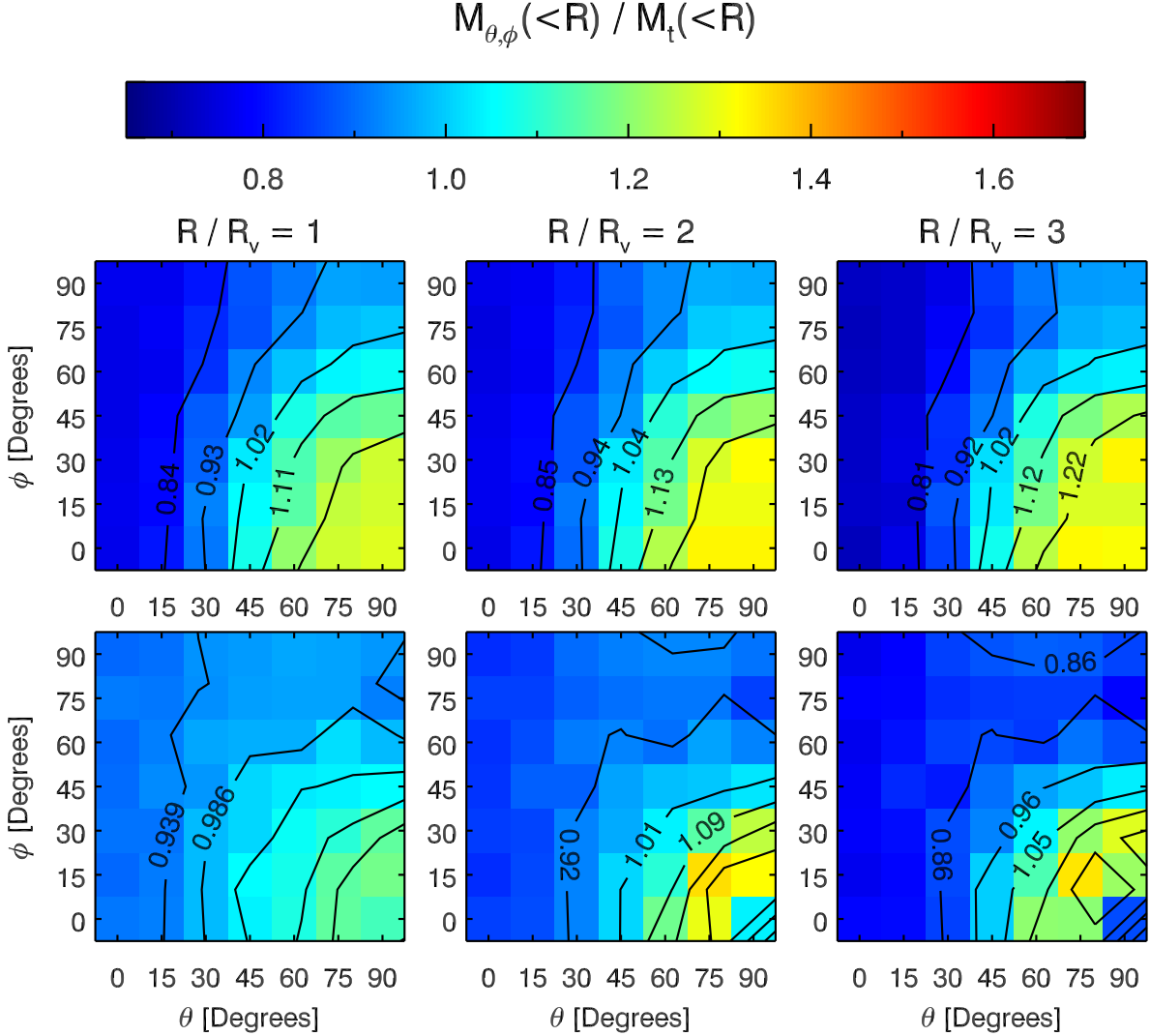}
		\vspace{-1.5cm}
		\caption{Effect of the cluster orientation on the mass estimate with the caustic method for the low mass bin i.e. $M_v \in \left[ 1,2 \right]\times 10^{14}\,h^{-1}M_{\sun}$. The panels show the mass estimates as a function of the orientation, relative to the true mass profile $M_t$. The three columns show results for three choices of radii. The top row shows mass estimates for the \textit{ellipsoidal stack} and the bottom row shows the same for the \textit{filamentary stack}. $\theta$ and $\phi$ indicates the line of sight in question, defined for each stack in Fig. \ref{fig:model}. For the top panels $(90^{\circ},0^{\circ})$ represents the sight line along the major axis, $(90^{\circ},90^{\circ})$ represents light line along minor axis and entire left side at $\theta=0^{\circ}$ represents the minor axis for any $\phi$. For the bottom panels $(90^{\circ},0^{\circ})$ represents the sight line along the filament, and entire left side ($\theta=0^{\circ}$ for any $\phi$) and top ($\phi=90^{\circ}$ for any $\theta$) represent sight lines orthogonal to the filament. The color for any $(\theta,\phi$) indicates the value of the mass estimate as indicated on the linear colorbar. The lines in each panel show equally-spaced isodensity contours.}
		\label{fig:bias_both}
	\end{center}
\end{figure*} 

\begin{figure*}
	\begin{center}
		\includegraphics[width=17	cm]{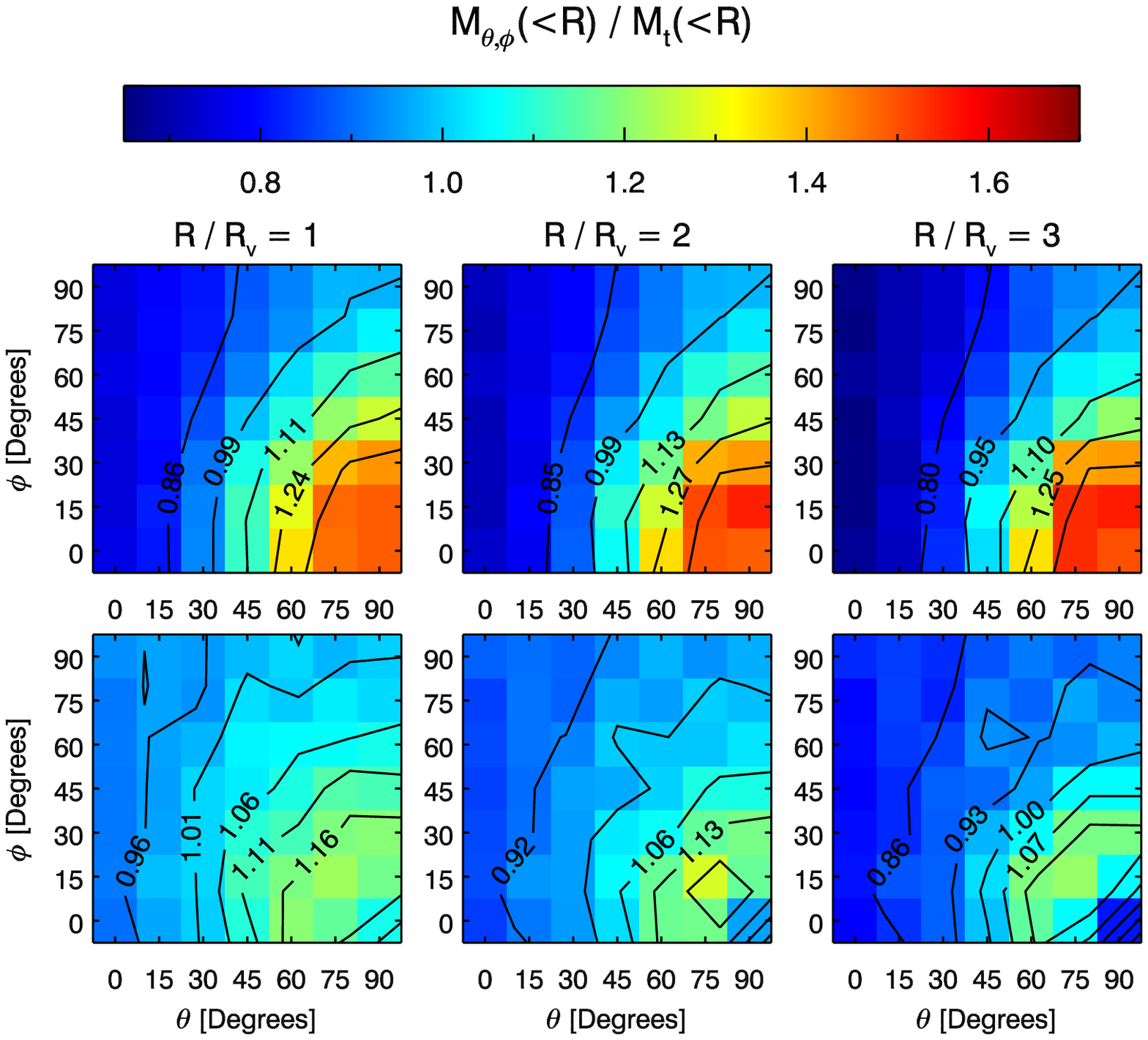}
		\vspace{-1.5cm}
		\caption{Same as Fig. \ref{fig:bias_both} but for the high mass bin i.e. $M_v \geq  2\times 10^{14}\,h^{-1}M_{\sun}$.}
		\label{fig:bias_both2}
	\end{center}
\end{figure*}

In order to gain a reference point for evaluating the anisotropic stacks, lines of sight where chosen to cover the half sphere of both the low and high mass bin \textit{spherical stack} in a $15^{\circ}\times 15^{\circ}$ grid from $\theta\in[0^{\circ},90^{\circ}]$ and $\phi\in[-180^{\circ},180^{\circ}]$. The mean cumulative mass profile $M_s(<R)$ was then calculated for the spherical stacks. This was done by choosing 10,000 pairs of $(\theta,\phi)$ randomly distributed on the sphere, interpolating the mass profiles from the $15^{\circ}\times15^{\circ}$ grid to obtain 10,000 mass profiles, and then taking the average profile. $M_s(<R)$ and its 95.4\% variability for the low mass stacks can be seen as the grey shaded area of Fig. \ref{fig:caustic_amplitudes} on the bottom. The black solid line shows the mean $M_t(<R)$ of the true cumulative mass profiles calculated from the full Bolshoi particle data set for each of the 230 clusters used in the three stacks. The caustic mass estimates of the \textit{spherical stack} agrees well with the true mass profile until 3$\,R_v$, from where the mass is underestimated. We note that this is the case given $\mathcal{F}_{\beta}=0.59$ and $\mathcal{F}_{\beta}=0.63$ for the two mass bins respectively, which motivates our choice. Through equation (\ref{eq:massprofile}) the caustic amplitude obtained from $M_s(<R)$ and its 95.4\% variability is shown in the top part of the same figure.

\begin{figure}
	\begin{center}
\hbox{\hspace{0.4cm}\includegraphics[width=8.8cm]{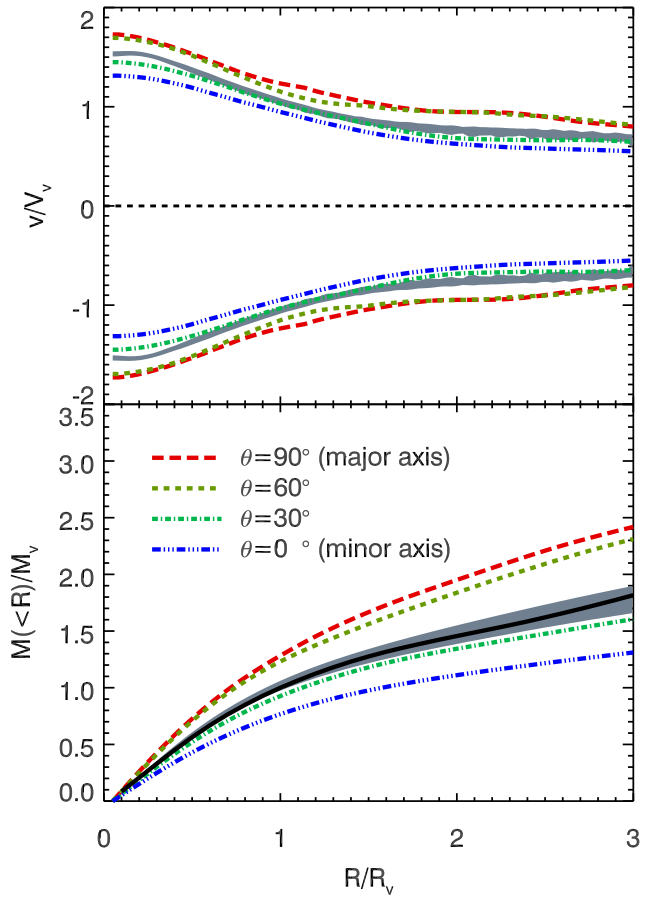}}
		\caption{\textit{Top:} Caustic amplitudes for the low mass bin \textit{ellipsoidal stack} using 4 different lines of sight with $\phi=0^{\circ}$ and $\theta$ as indicated by the legend. This corresponds to moving gradually from a line of sight along the major axis of the stack towards a line of sight along the minor axis. The grey area shows two standard deviations of caustic amplitudes through the spherical stack, as obtained by integrating the grey area in the bottom panel of this figure (see below). \textit{Bottom}: The resultant cumulative mass profiles as obtained by applying equation (\ref{eq:massprofile}) to the amplitudes of the top plot. Grey area shows two standard deviations of 10,000 mass profiles from random lines of sight through the \textit{spherical stack}. The solid black line shows the true mean mass profile of the 230 clusters as calculated from the full Bolshoi particle data set.}
		\label{fig:caustic_amplitudes}
	\end{center}
\end{figure}

To fully map out the angular dependency of the mass estimates of the caustic technique on the anisotropic stacks, lines of sight where chosen in the same $15^{\circ}\times 15^{\circ}$ grid as above, covering the half sphere. From an ellipsoidal point of view, the four octants that span the half sphere are equivalent. Thus properties like position, velocity dispersion and inferred caustic mass for the four example lines of sight 
$(\theta,\phi )=\left\lbrace 
(-165^{\circ},15^{\circ}),
(-15^{\circ},15^{\circ}),
(15^{\circ},15^{\circ}),
(165^{\circ},15^{\circ}) \right\rbrace$  should be symmetric in these octants. By using the chosen grid, the cumulative mass profiles $M_{\theta,\phi}(<R)$ for each octant were calculated, and the average octant was taken. The resulting profiles were normalized to the true mass profile $M_t(<R)$. Because the mass profiles of the \textit{spherical stack} $M_s(<R)$ and the true mass profile $M_t(<R)$ agree so well, the normalized profiles express an intrinsic bias of the caustic method from spatial anisotropy, relative to the caustic mass profile inferred when the assumption of sphericity is perfectly valid.

The three top panels of Fig. \ref{fig:bias_both} show the relative cumulative mass estimates at 1, 2 and 3$\,R_{v}$ for the average octant of the low mass bin \textit{ellipsoidal stack}. The lower right corner of each panel at $(90^{\circ},0^{\circ})$ represents the line of sight along the major axis. The upper right corner at $(90^{\circ},90^{\circ})$ represents the intermediate axis and the entire left side at $\theta=0^{\circ}$ represents the minor axis for any $\phi$. The color of the figure shows the value of the cumulative mass at a given radius relative to the spherical mass estimate. Table \ref{tab:ellipsoid} of the Appendix shows the values for each set of angles. The bottom three panels of Fig. \ref{fig:bias_both} show the same as the top three but for the \textit{filamentary stack}. The lower right corner of each panel represents the line of sight along the maximal density direction i.e. along the filament. The entire upper ($\phi=90^{\circ}$) and left ($\theta=0^{\circ}$) side represents lines of sight orthogonal to the filament. Table \ref{tab:filament} of the Appendix shows the values for each set of angles in each panel. 

We stress that the masses displayed here (and throughout the paper) are evaluated at multiples of the true virial radius $R_v$, not the observed virial radii $R_{v,obs}$ provided by the caustic method. Upon iteratively determining the caustic mass and caustic virial radius $R_{v,obs}$, the masses at 1,2 and 3 $R_{v,obs}$ deviate slightly from those displayed in this figure. As an example the mass profiles of Fig. \ref{fig:bias_both} are evaluated at $R_{v,obs}$ rather than $R_v$ in Fig. \ref{fig:obsrvir} of the Appendix.

Fig. \ref{fig:bias_both2} also shows the angular dependency of the caustic mass estimates for the \textit{ellipsoidal} and \textit{filamentary stack} but for the high mass bin i.e. the same thing as Fig. \ref{fig:bias_both} but for more massive clusters. Tables \ref{tab:ellipsoidMB2} and \ref{tab:filamentMB2} of the Appendix show the values for each set of angles in each panel.

Table \ref{tab:results} shows values derived from the data displayed in Fig. \ref{fig:bias_both} and \ref{fig:bias_both2}. The 'Mean mass' columns were generated in the same way as the spherical mean mass profiles $M_s(<R)$ by choosing 10,000 sets of ($\theta$,$\phi$) randomly distributed on the sphere, and then interpolating the mass estimates shown in Fig. \ref{fig:bias_both} and \ref{fig:bias_both2}. The mean of these 10,000 mass estimates are then the 'Mean mass' at a given radius, but normalized to $M_t(<R)$. The 'Scatter in masses' columns show the 68.3\% standard deviation of the normalized 10,000 mass estimates for each stack and radius. Finally the 'Maj/min mass' columns show for the for the \textit{ellipsoidal stack} the major axis mass estimate relative to the minor axis mass estimate for each of the top panels in Fig. \ref{fig:bias_both} and \ref{fig:bias_both2} i.e. at 1, 2 and 3$\,R_v$. For the \textit{filamentary stack} the columns show the mean of the mass estimates at 
$(\theta,\phi)=
\{
\left(60^{\circ},0^{\circ}\right), 
\left(60^{\circ},15^{\circ}\right), $ $
\left(75^{\circ},0^{\circ}\right), 
\left(75^{\circ},15^{\circ}\right),
\left(75^{\circ},30^{\circ}\right),
\left(90^{\circ},15^{\circ}\right),
\left(90^{\circ},30^{\circ}\right)
\}$ 
relative to the mean of the seven mass estimates orthogonal to the filament at $\phi=90^{\circ}$. This represents the largest mass estimate relative to the lowest, taking into account the symmetry of the stack because the biggest mass overestimate appears when the line of sight is tilted slightly at $\sim20^{\circ}$ to the filament axis. This will be discussed in detail in the next section.

Fig.~\ref{fig:bias_both} and Fig.~\ref{fig:bias_both2}, and the 'Maj/min mass' and 'Scatter in masses' columns in Table~\ref{tab:results} describe a mass bias subject to known cluster orientation with respect to the sight line, e.g. as inferred from lensing observations \citep[see e.g.][]{Limousin2013}. This needs to be distinguished from a bias of an average mass obtained from an ensemble of measurements when clusters are aligned randomly with respect to the sight line. This bias is described by the 'Mean mass' column.

\section{Discussion}\label{sec:discussion}
	The increased statistics of the two \textit{spherical stacks} allow for a very consistent estimate of their true caustic amplitudes and masses, regardless of the choice of line of sight. Each of the \textit{spherical}, \textit{ellipsoidal} and \textit{filamentary stacks} consists of the same number of clusters within each mass bin, and thus have the same true cumulative mass profile. Any cumulative mass profiles found in the \textit{ellipsoidal} and \textit{filamentary stacks} that differ from those of the \textit{spherical stacks} therefore result from a spatial anisotropy bias in the caustic technique.
Fig. \ref{fig:caustic_amplitudes} shows clearly how the mass profiles from the caustic procedure can be affected by asphericity for the low mass bin: When observing the \textit{ellipsoidal stack} along its most elongated direction (major axis at $(\theta,\phi)=(90^{\circ},0^{\circ})$) the caustic amplitude and its subsequent mass estimate (red long dashed curves) are larger than for a spherical object of the same true mass (grey areas). By choosing lower $\theta$ for the line of sight and thereby making the cluster less elongated in these directions, the caustic amplitudes systematically drop in magnitude until they reach a minimum at $\theta=0^{\circ}$ (blue dash-dot-dot-dotted curves), well below the mass estimate of the spherical object. Clearly the more an elongated object has its axis of most elongation along the line of sight, the larger cumulative mass the caustic procedure will find at any radius. Similarly, the closer its least elongated direction is to the line of sight, the smaller cumulative mass the caustic procedure will find.

\begin{table*}\begin{center}
    \caption{Values of triaxiality and mass estimates for the different stacks under consideration. The three 'Triaxiality' columns show the intermediate to major axis ratio $b/a$, the minor to major axis ratio $c/a$ and the triaxiality parameter $T$ for each of the stacks in questions for particles within $R_v$. The 'Mean mass' columns show the mean of 10,000 mass measurements obtained from 10,000 interpolations of each panel in Fig. \ref{fig:bias_both} and Fig. \ref{fig:bias_both2} for 10,000 directions randomly distributed on the sphere. The 'Scatter in masses' columns show the scatter defined as the standard deviation of the 10,000 mass measurements. The 'Maj/min mass' columns show the ratio of the mass estimate obtained when observing along the major axis to the mass estimate obtained when observing along the minor axis. All mass values are normalised by the true cluster mass $M_t(<R)$.\label{tab:results}}
    \begin{tabular}{ccccccccccccc}
    \toprule
    Stack name	& \multicolumn{3}{c}{Triaxiality} & \multicolumn{3}{c}{Mean mass} & \multicolumn{3}{c}{Scatter in masses} & \multicolumn{3}{c}{Maj/min mass} \\
     & $b/a$ & $c/a$ & $T$ & $1\,R_v$ & $2\,R_v$ & $3\,R_v$ & $1\,R_v$ & $2\,R_v$ & $3\,R_v$ & $1\,R_v$ & $2\,R_v$ & $3\,R_v$\\
    \hline \\   \vspace{2mm}	
    $M_v \in \left[ 1,2 \right]\times 10^{14}\,h^{-1}M_{\sun}$ \\
    Ellipsoidal Stack 	&	0.78 & 0.65 	& 0.69 	& 1.02 &	1.04	& 1.02	&	0.14	&	0.16	&	0.17	&	1.68	&	1.74	&	1.82	\\
    \vspace{2mm}
    Filamentary Stack	&	0.89	& 0.86	& 0.81 	& 1.01 & 0.99	& 0.96 	&	0.07&	0.12	&	0.14	&	1.21	&	1.38	&	1.45	\\ 

     $M_v \geq  2\times 10^{14}\,h^{-1}M_{\sun}$ & \\
    Ellipsoidal Stack 	& 	0.76	& 	0.63	&	0.70	& 1.08 &1.06 	& 1.03	&	0.20	& 0.22	&0.24	&	1.98&	2.07	&	2.25	\\
    \vspace{2mm}
    Filamentary Stack	&	0.88	&	0.85	&	0.81	&	1.07	&	1.02	& 	0.98& 0.08	& 0.09	& 0.10	&	1.22	&	1.28	& 1.31	\\ 

    \hline
    \end{tabular}
\end{center}\end{table*}

\subsection{Low mass bin}

Focusing on the low mass bin, Fig. \ref{fig:bias_both} shows the mass estimates at 1, 2 and 3$\,R_{v}$ for a line of sight at varying angles $\theta$ and $\phi$ for both the low mass bin \textit{ellipsoidal} and \textit{filamentary stack}. All mass estimates are normalized by the true cumulative mass profile $M_t(<R)$. Consistent with expectations from Fig. \ref{fig:caustic_amplitudes} the largest deviations from spherical mass estimates in the \textit{ellipsoidal stack} (top three panels) appears once the line of sight coincides with either the major ($(\theta,\phi)=(90^{\circ},0^{\circ})$) or the minor axis ($\theta = 0^{\circ}$ for any $\phi$). The mass estimate along the intermediate axis ($(\theta,\phi)=(90^{\circ},90^{\circ})$) lies comfortably in between the two extremes and generally the mass estimates follow a smooth transition between the extremes of the three principal axes. Along the major axis the cumulative mass estimates are 1.28, 1.32 and 1.31 times larger than the spherical mass estimates at 1, 2 and 3$\,R_{v}$ respectively. Along the minor axis the cumulative mass estimates are a factor of 0.77, 0.76 and 0.72 lower than the spherical mass estimates at the same radii as before. Taking the ratio of the maximal mass estimate to the minimal mass estimate, this effect spans a factor of 1.68, 1.74 and 1.82 at 1, 2 and 3$\,R_{v}$ respectively, as noted in Table \ref{tab:results}. 
Table \ref{tab:results} also lists the mean mass of 10,000 random direction interpolations of the three top panels, i.e. an estimate of the bias. For the \textit{ellipsoidal stack} this is close to one at any of the measured radii, meaning that caustics used on ellipsoidal structure on average is unbiased relative to spherical cluster mass estimates. The standard deviation of the 10,000 random direction mass profiles relative to the true mass is also listed in Table \ref{tab:results}. The listed values of 0.14 at 1 $R_v$ and 0.17 at $3\,R_v$ is in good agreement with the lower limits of \cite{serra2011} and \cite{gifford2013_1} who both estimated scatter for a variable number of member galaxies.

The \textit{filamentary stacks} are different from the \textit{ellipsoidal stacks} both by construction and in the resulting bias effect. Inspecting Fig. \ref{fig:bias_both} for the low mass bin \textit{filamentary stack} (bottom three panels) one sees a general increase in mass at angles close to $(\theta,\phi)=(90^{\circ},0^{\circ})$ which represents a line of sight directly along the filament, and a general decrease in mass at line of sight orthogonal to the filament. Upon closer inspection of the maximal mass estimate in the lower right corners of the three panels, one notes that the maximum occurs as the line of sight is slightly tilted to $\sim20^{\circ}$ with respect to the filament axis i.e. at 
$(\theta,\phi)=
\{
\left(60^{\circ},0^{\circ}\right), 
\left(60^{\circ},15^{\circ}\right), $ $
\left(75^{\circ},0^{\circ}\right), 
\left(75^{\circ},15^{\circ}\right),
\left(75^{\circ},30^{\circ}\right),
\left(90^{\circ},15^{\circ}\right),
\left(90^{\circ},30^{\circ}\right)
\}$ . The mean of these estimates at this tilt is a factor of 1.13, 1.23 and 1.23 larger than $M_t(<R)$ at 1, 2 and 3$\,R_{v}$ respectively. Whereas the filament direction was chosen uniquely as the maximal density direction, the two other directions were selected at random within the plane orthogonal to the filament. As such, a different realization of the stack would yield a different orientation of all the clusters two unit vectors within the orthogonal plane. Indeed slight differences were observed for different realizations of the stack, and the column represented by $\phi=90$ represents 7 lines of sight within the plane orthogonal to the filament, and thus we would expect the same mass estimate. We therefore take the average of these seven values and obtain a mass estimate of the \textit{filamentary stack} lower by factor of 0.93, 0.89 and 0.84 at 1, 2 and 3$\,R_{v}$ when observing orthogonal to the line of sight. This yields a maximum-to-minimum ratio of 1.21, 1.38 and 1.45 at these radii. As such, the effect of filamentary anisotropy is slightly smaller than, yet comparable to the effect of elongation of the central structure for the low mass bin. From Table \ref{tab:results} we see that the mean mass is close to the true mass of the clusters than the spherical case, while the scatter in masses is slightly smaller for the case of filaments than for the case of elongation.

In order to explain the shift in maximum of mass estimates for the low mass \textit{filamentary stack} of $\sim 20^{\circ}$ we remind that the \textit{filamentary stack} is constructed by orienting individual structures after the maximal density direction as found within a search cone frustum spanning 30$^{\circ}$ in angle between 1 and 5$\,R_{v}$ (see Section \ref{sec:filament_geo}). As can be seen in e.g. \cite{cuesta2008}, the radial velocity component of particles in a galaxy cluster is zero on average up to $\sim 1.5\,R_{v}$. Further out between $\sim 1.5\,R_{v}$ and $3.5\,R_{v}$ the radial velocity tends to be negative in a zone of infall towards the cluster. Still further out the Hubble expansion takes over and the radial velocity becomes positive and increases with radius (see Fig. 4 of \cite{cuesta2008}). The cone frustum covers roughly equally much of the infall zone and the Hubble expansion zone, and as such we expect the particles within the filament to appear more or less symmetrically around $v=0$ in ($R$-$v$)-phase space. 
\begin{figure*}
    \vspace{-2.1cm}
	\hspace{-2cm}
		\hbox{\hspace{0.4cm}\includegraphics[width=17cm]{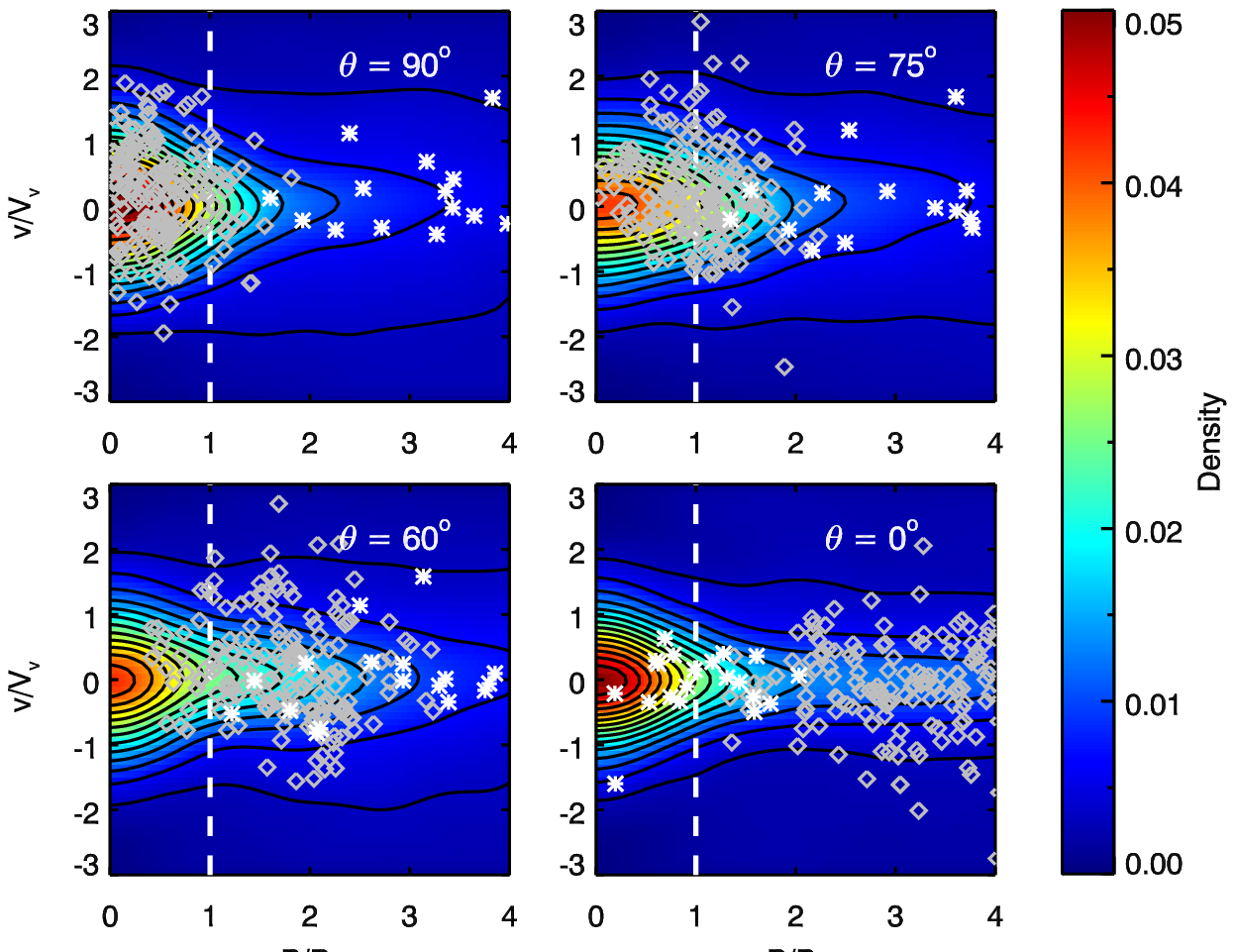}}
	\vspace{0.8cm}
		\caption{Location of filament particles in ($R,v$) phase-space using different projections. The data is from the low mass bin \textit{filamentary stack} (i.e. $M_v \in \left[ 1,2 \right]\times 10^{14}\,h^{-1}M_{\sun}$) obtained by a constant $\phi = 0^{\circ}$ and a $\theta$ as indicated in white in each panel. $\theta = 90^{\circ}$ indicates a sight line directly along the filament, and $\theta =0^{\circ}$ indicates the sight line orthogonal to the filament. The coloured contours represent the density obtained by the kernel density estimation method. The grey diamonds and white stars are projections of a total of 400 randomly drawn particles from within the filament frustum (grey diamonds) or a similar volume orthogonal to the filament (white stars). The white dashed line marks the location of $R_v$. Note that some points may fall outside the plotting range. Also note, that even though the phase space density diagrams have been corrected for the two-halo term, the number of particles displayed has not. }
		\label{fig:trace}
\end{figure*}
Fig. \ref{fig:trace} shows the phase space density distribution for 4 lines of sight through the low mass bin \textit{filamentary stack}. It is clear from the 4 panels that the density distribution changes as the sight-line is shifted. Particles within the filament are the main cause of the $\sim 20^{\circ}$ shift that yields the highest caustic mass estimate, and to see why this is we consider two volumes: The cone frustum defining the filament (enclosing filament particles), and a similar volume orthogonal to the filament along the positive $y$-axis (enclosing non-filament particles). From the joint set of particles enclosed by the two volumes, we draw a random subset of 400 particles. Points inside the filament frustum are indicated as grey diamonds in Fig. \ref{fig:trace}, those in the orthogonal volume are shown as white stars. There are many more grey diamonds than white stars simply because there are many more filament particles than non-filament particles. The placement of the filament particles in projection therefore affects phase-space strongly. The important thing to note is where filament particles place themselves in projection for a given line of sight. When viewing the stack along its filament at $\theta =90^{\circ}$ most of the filament particles project well within $R_{v}$. Here they add some, but not much to the overall mean velocity dispersion which is used to calibrate the derived caustic amplitude (see equation (\ref{eq:calibration})). When viewing the filament at a slight angle with $\theta = 75^{\circ}$ the filament particles spreads over the entire inner part of the cluster in projection, acting to perturb the isodensity contours and increase the velocity dispersion. This in turn results in an overall larger caustic amplitude and thus larger mass estimate. As the line of sight is shifted even more to $\theta = 60^{\circ}$, the filament particles have left the inner parts of the cluster in projection, and thus affect the caustic amplitude chosen by equation (\ref{eq:calibration}) less. Finally at line of sight orthogonal to the filament at $\theta=0^{\circ}$ all the filament particles are projected to high radii, and interfere only with the mass estimate at these radii. 

Fig. \ref{fig:obsrvir} of the Appendix shows the same as Fig. \ref{fig:bias_both}, however with masses evaluated at observed $R_{v,obs}$ rather than the true $R_v$. The caustic method systematically overestimates the virial radius with line of sight along the semi-major axis, and underestimates it along the semi-minor axis. Thus the mass evaluated at the observed radii will depend on line of sight in the same way as in Fig. \ref{fig:bias_both} as can be seen in the resemblance of the two figures, but with a slightly larger spread for $R_{v,obs}$.

\subsection{High mass bin}

Upon comparing Fig. \ref{fig:bias_both} for the low mass bin with Fig. \ref{fig:bias_both2} for the high mass bin, one notices some resemblance. For the high mass bin \textit{ellipsoidal stack} represented in the three top panels of Fig. \ref{fig:bias_both2} there is still a systematic effect of overestimation of mass along the major axis and an underestimation along the minor axis, with the intermediate axis mass estimate lying in between. The effect is even larger for the high mass bin than low mass bin, with a major to minor axis mass ratio of 1.98, 2.07 and 2.25 at 1, 2 and $3\,R_v$ respectively. As can be seen in Table \ref{tab:results} the low and high mass \textit{ellipsoidal stacks} are very similar in respect to triaxiality and elongation, so the larger caustic mass estimates may therefore result from differences in velocity distribution or differences in elongation evaluated from particles at $>R_v$. The 'Mean mass' columns of Table \ref{tab:results} show that elongation of the cluster imposes a $<5\%$ systematic bias in the mean of mass profiles on average within $R_v$. The 'Scatter in masses' column show the slightly higher scatter in comparison with the low mass bin, however still in line with expectations.

For the high mass \textit{filamentary stack} shown in the three bottom panels of Fig. \ref{fig:bias_both2} there is still a maximum overestimation at a $\sim$20$^{\circ}$ tilt from line of sight along the filament. Where ellipsoidal effects caused a larger scatter in mass estimates for the high mass bin, the effect of filaments seems to be the same if not marginally reduced. As such the \textit{filamentary stack} showed a maximum to minimum mass estimate ratio of only 1.22, 1.28 and 1.31 at 1, 2 and 3$\,R_v$. \cite{colberg2005} reports that the number of filaments associated with a cluster increases for increasing halo mass, which may act to smear out the contrast between line of sight along the filament, and line of sight across the filament in the stack. Furthermore a large cluster does not necessarily connect to a large filament, and as such the heavier clusters are less affected by the filament.

\subsection{Triaxiality}
It is interesting to quantify whether the mass measurements shown in Fig. \ref{fig:bias_both} and \ref{fig:bias_both2} for the \textit{ellipsoidal stacks} depend on triaxiality, i.e. if the distribution of mass estimates in $(\theta,\phi)$-space looks different for mainly oblate or prolate clusters. By splitting the \textit{ellipsoidal stacks} into substacks of the most prolate and the most oblate, we obtain an 'oblate' stack with  $T=0.31$ using the 40 most oblate clusters and a 'prolate' stack with $T=0.89$ using the 40 most prolate clusters for the low mass bin. Similarly we make a 19 cluster oblate ($T=0.34$) and prolate ($T=0.89$) substack for the high mass bin. We perform the same angular gridded anlysis of all substacks as for the full stacks. Fig. \ref{fig:bias_obpro_low} of the Appendix shows the mass estimates of the low mass bin prolate (top panels) and oblate (bottom panels) substack relative to the true mass profile $M_t(<R)$\footnote{Note that $M_t(<R)$ is the mean profile for the full stacks, and as such may differ slightly from the true mass profile of the prolate and oblate substacks}. Upon comparing the substacks of Fig. \ref{fig:bias_obpro_low} to the \textit{ellipsoidal stack} in the top panels of Fig. \ref{fig:bias_both} one sees very little variation from the low mass bin \textit{ellipsoidal stacks} to the oblate and prolate conterparts. Similarly the high mass bin shows little variation between the oblate and prolate substacks of Fig. \ref{fig:bias_obpro_high} of the Appendix and the high mass bin \textit{ellipsoidal stacks} in Fig. \ref{fig:bias_both2}. For this sparse sample of galaxy clusters we therefore conclude that oblateness and prolateness affect mass estimation in more or less the same manner. Naturally by cutting the stack cluster population the statistics suffer, and the estimates of Figs. \ref{fig:bias_obpro_low} and \ref{fig:bias_obpro_high} of the Appendix should be considered less certain. All values for the prolate and oblate substacks are summarized in Table \ref{tab:results_oblate_prolate} of the Appendix.
\section{Conclusion}\label{sec:conclusion}
	We studied the bias in the mass estimate of galaxy clusters based on the caustic technique, resulting from orientation of clusters with respect to the line of sight. We analysed dark matter particle data from the Bolshoi $N$-body simulation for a set of 230 dark matter halos at $M_v \in \left[ 1,2 \right]\times 10^{14}M_{\sun}h^{-1}$ and a set of 101 dark matter halos at $M_v > 2\times 10^{14}\,M_{\sun}h^{-1}$. Each of the halos were superposed concentrically in 3 separate stacks differing only by orientation of the individual halos: The \textit{ellipsoidal stack} had each halo rotated such that the three principal axes from its shape tensor inside $R_{v}$ were aligned with the $x$-, $y$- and $z$-axis. The \textit{filamentary stack} had each halo oriented after the direction of maximal density inside a cone frustum of angle 30$^{\circ}$ between $1\,R_{v}$ and $5\,R_{v}$, such that they were all oriented along the positive $x$-axis. The \textit{spherical stack} had all halos stacked with completely random orientations, where each halo was used 10 times with 10 random orientations to increase the sphericity of the stack.
Using the now standard caustic technique for mass estimation of galaxy clusters we projected each of the stacks to a $15^{\circ}\times15^{\circ}$ angular grid in both mass bins and estimated the apparent caustic amplitude and cumulative mass profile for all angles. When using the mass estimate of the \textit{spherical stack} for the low mass bin we found a good correspondence with the true cumulative mass profiles when using $\mathcal{F}_{\beta} = 0.59$, as shown in Fig. \ref{fig:caustic_amplitudes}. For the high mass bin we used $\mathcal{F}_{\beta} = 0.63$. Using the \textit{spherical stack} as reference for the idealized situation under which the caustic method can be applied, we saw that the caustic amplitudes varied systematically with line of sight in the \textit{ellipsoidal stack}. Using a line of sight along the major axis, we found that the caustic mass estimate was overestimated by a factor of 1.28, 1.32 and 1.31 relative to the true mass at 1, 2 and 3$\,R_{v}$ respectively. 
Similarly with a line of sight along the minor axis the mass was underestimated with a factor of 0.77, 0.76 and 0.72 at the same radii relative to true mass. Taking the ratio of the maximal mass estimate to the minimal mass shows that the effect is as large as a factor of 1.68, 1.74 and 1.82 at 1, 2 and 3$\,R_{v}$ respectively. We found that on average the caustic mass estimates for the \textit{ellipsoidal stack} were unbiased relative to the true mass. 
For the low mass bin \textit{filamentary stack} the same analysis was performed and yielded an overestimation of cumulative mass of 1.13, 1.23 and 1.23 at 1, 2 and 3$\,R_{v}$ with line of sight slightly tilted to along the filament, and an underestimation by factor of 0.93, 0.89 and 0.84 at 1, 2 and 3$\,R_{v}$ when observing orthogonal to the filament. This gave a maximum to minimum ratio of 1.21, 1.38 and 1.45 at these radii, which is somewhat smaller than, but comparable to the ratios of the \textit{ellipsoidal stack}. 
We investigated the fact that the largest mass overestimate occurred when the line of sight was tilted at a $\sim 20^{\circ}$ angle to the filament. We traced this effect to the location of the filament particles in projection, which had a maximal influence on velocity dispersion and caustic isodensity contours at $\sim 20^{\circ}$ as demonstrated in Fig. \ref{fig:trace}. 

For the large of the two mass bins, the systematics of the effects were the same, but the magnitude larger. The caustic masses were still overestimated along the major axis of the \textit{ellipsoidal stack} and underestimated along the minor axis. This effect was however even more significant for the high mass bin with a maximum mass to minimum mass ratio of 1.98, 2.07 and 2.25 at 1, 2 and $3\,R_v$ respectively. On the other hand the \textit{filamentary stack} showed a slightly lower influence of filaments on larger clusters, spanning only a maximum to minimum mass ratio of 1.22, 1.28 and 1.31 at 1, 2 and $3\,R_v$.

We found that the scatter in mass estimates due to ellipsoidal and filamentary anisotropy was in good agreement with the lower limits of \cite{serra2011} and \cite{gifford2013_1} who both estimated scatter in caustic mass for a variable number of member galaxies. Thus a significant portion of the scatter presented in these references may be explained by anisotropic models in the form of ellipsoids, and to a lesser extent filaments.

To test the sensitivity of caustic mass estimates on oblate- and prolateness we considered substacks of the 40 most oblate and prolate halos of the low mass bin \textit{ellipsoidal stack} as defined by $T$, and substacks of the 19 most oblate and prolate halos of the high mass bin \textit{ellipsoidal stack}. We found no significant variation by comparing oblate to prolate mass estimates, or by comparing the oblate and prolate masses to the full \textit{ellipsoidal stack}.

The mass estimates plotted it Fig. \ref{fig:bias_both} and \ref{fig:bias_both2} can be used to correct caustic mass estimates when cluster orientation is known. The values from these figures are available in Table \ref{tab:ellipsoid}, \ref{tab:filament}, \ref{tab:ellipsoidMB2} and \ref{tab:filamentMB2} of the Appendix. 

The caustic method of mass estimation performs well when the condition of cluster sphericity is met. If however spatial anisotropy is present in the form of cluster elongation of filamentary structure, the caustic masses are strongly dependent on the line of sight through the cluster. Even within the virial radius the mass estimates may vary by a factor of $\sim2$ for high-mass clusters, and as such great care should be taken when applying this method.

The reason why the caustic mass measurements depend on the cluster orientation with respect to the line of sight is the anisotropy of the spatial as well as velocity distribution of galaxies in clusters. Considering the velocity component, this means that the effect of asphericity on the measurement of cluster masses is expected to be a generic feature of all kinematical methods for the cluster mass determination, e.g. methods based on the virial theorem or the scaling relation between cluster mass and the line-of-sight velocity dispersion \citep{Biviano2006,Saro2013}, methods based on the Jeans analysis of the velocity moments profiles \citep{Sanchis2004,Lokas2006}, methods using models of the projected phase space distribution \citep{Wojtak2009,Mamon2013} or dynamical models of the infall velocity profile \citep{Falco2014} \citep[for a comparison between a broad range of available methods see][]{Old2014}. The discrepancy between the measured and the actual cluster mass may differ between the methods; therefore, our results cannot be regarded as a general prediction for all of them. However, the substantial mass discrepancy shown for the caustic technique provides strong motivation for detailed studies of this effect in all other kinematical methods.
\section*{Acknowledgments}
The authors would like to thank Daniel Gifford and Doron Lemze for helpfully answering questions regarding the caustic technique and useful comments. We also thank the anonymous referee for constructive comments which helped to improve the work. The Dark Cosmology Centre is funded by the Danish National Research Foundation. RW acknowledges support through the Porat Postdoctoral Fellowship.
\bibliographystyle{mn2e}

\newpage
\clearpage
\appendix

\begin{figure*}
	\begin{center}
		\includegraphics[width=17	cm]{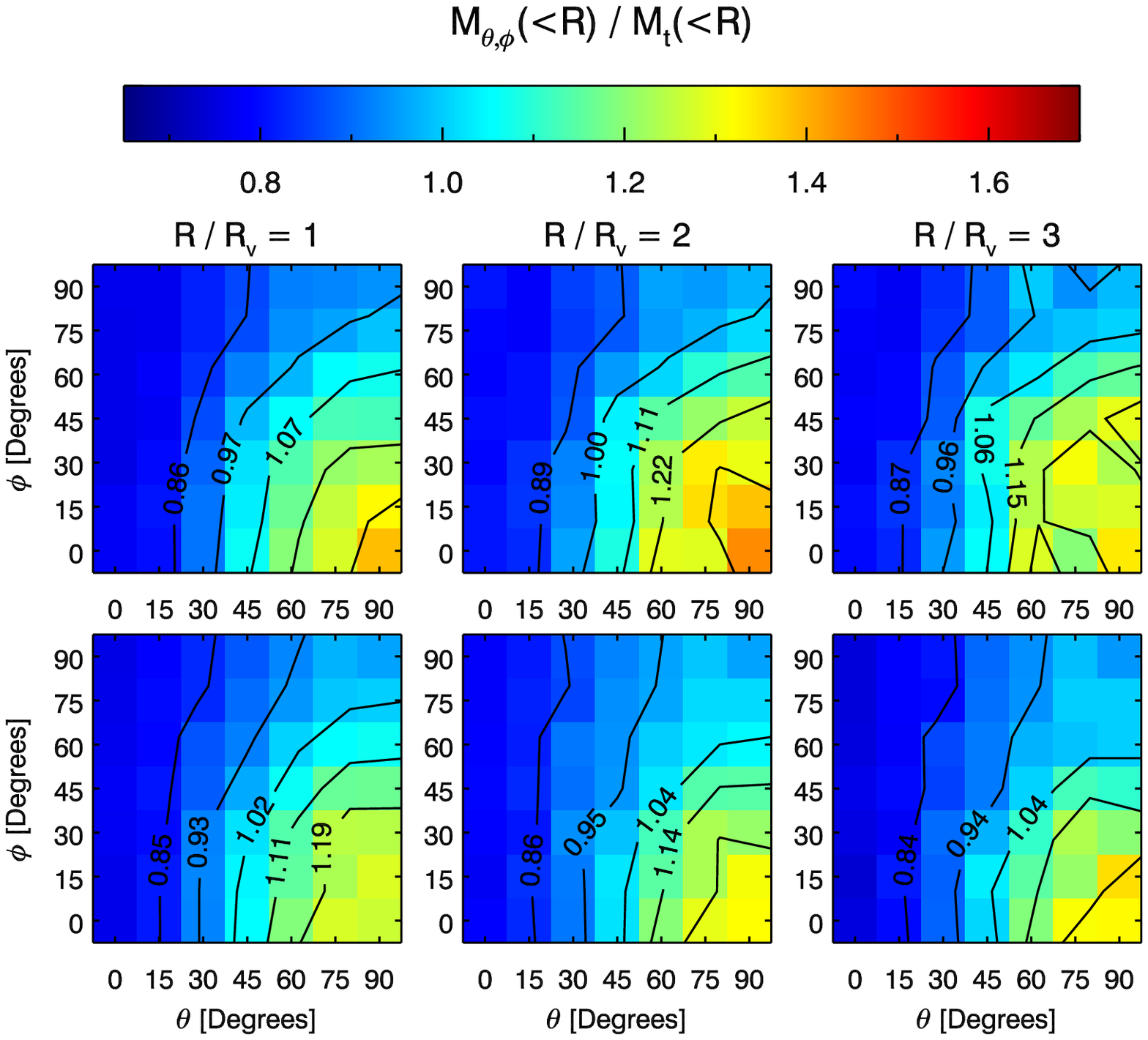}
		\caption{Effect of the cluster orientation on the caustic mass estimate for the 40 most oblate (top panels) and prolate (bottom panels) clusters as sorted by triaxiality $T$ for the low mass bin. The panels show the mass estimates as a function of the orientation, relative to the true mass profile $M_t(<R)$. The three columns show results for three choices of radii. $\theta$ and $\phi$ indicate the line of sight in question, defined for each stack in Fig. \ref{fig:model}. For the top panels $(90^{\circ},0^{\circ})$ represents the sight line along the major axis, $(90^{\circ},90^{\circ})$ represents light line along minor axis and entire left side at $\theta=0^{\circ}$ represents the minor axis for any $\phi$. The color for any $(\theta,\phi$) indicates the value of the mass estimate as indicated on the linear colorbar. The lines in each panel show equally-spaced isodensity contours. }
		\label{fig:bias_obpro_low}
	\end{center}
\end{figure*}

\begin{figure*}
	\begin{center}
		\includegraphics[width=17	cm]{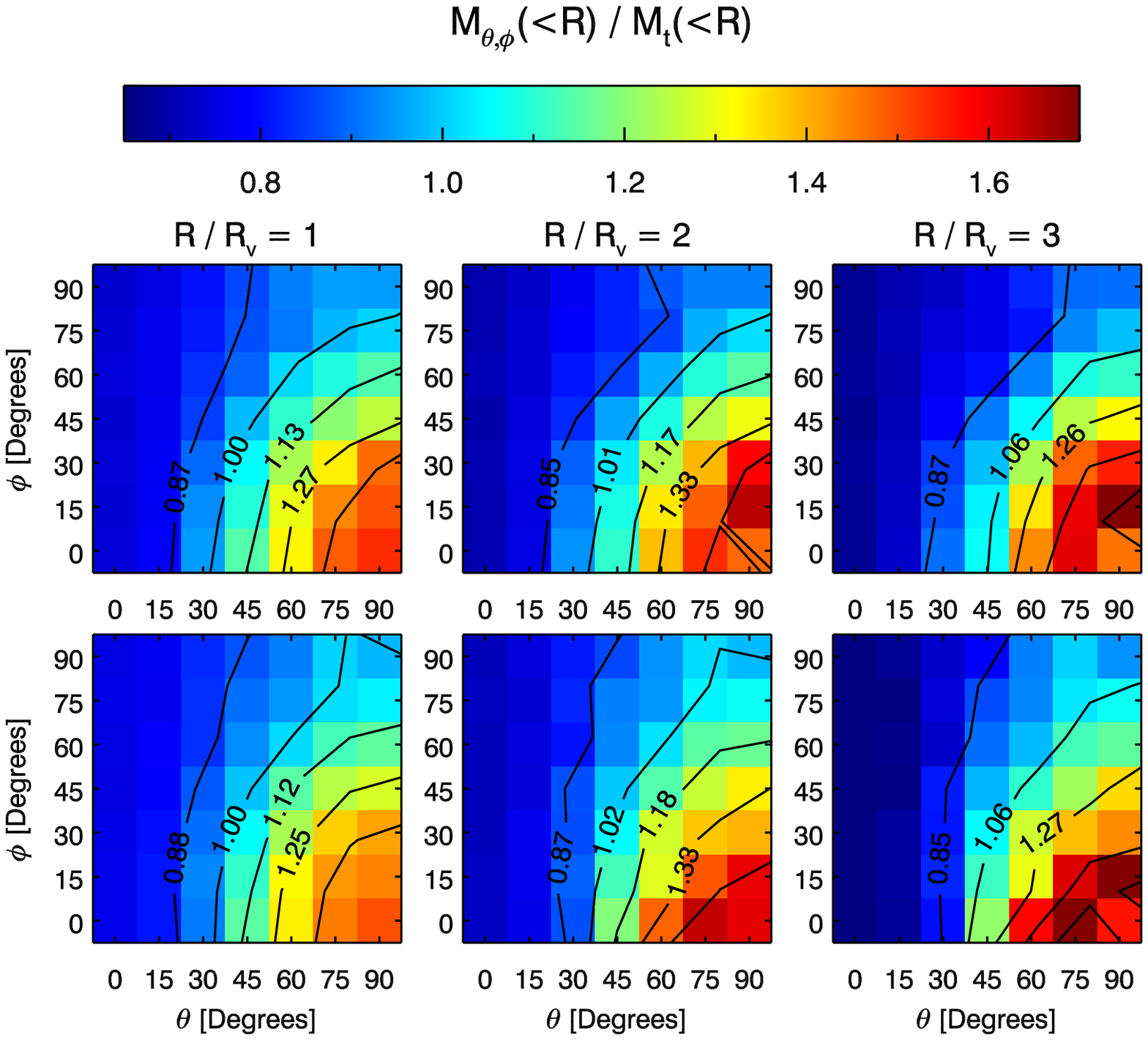}
		\caption{Effect of the cluster orientation on the caustic mass estimate for the 19 most oblate (top panels) and prolate (bottom panels) clusters as sorted by triaxiality $T$ for the high mass bin. The panels show the mass estimates as a function of the orientation, relative to the true mass profile $M_t(<R)$. The three columns show results for three choices of radii. $\theta$ and $\phi$ indicate the line of sight in question, defined for each stack in Fig. \ref{fig:model}. For the top panels $(90^{\circ},0^{\circ})$ represents the sight line along the major axis, $(90^{\circ},90^{\circ})$ represents light line along minor axis and entire left side at $\theta=0^{\circ}$ represents the minor axis for any $\phi$. The color for any $(\theta,\phi$) indicates the value of the mass estimate as indicated on the linear colorbar. The lines in each panel show equally-spaced isodensity contours.}
		\label{fig:bias_obpro_high}
	\end{center}
\end{figure*}

\begin{figure*}
	\begin{center}
		\includegraphics[width=17	cm]{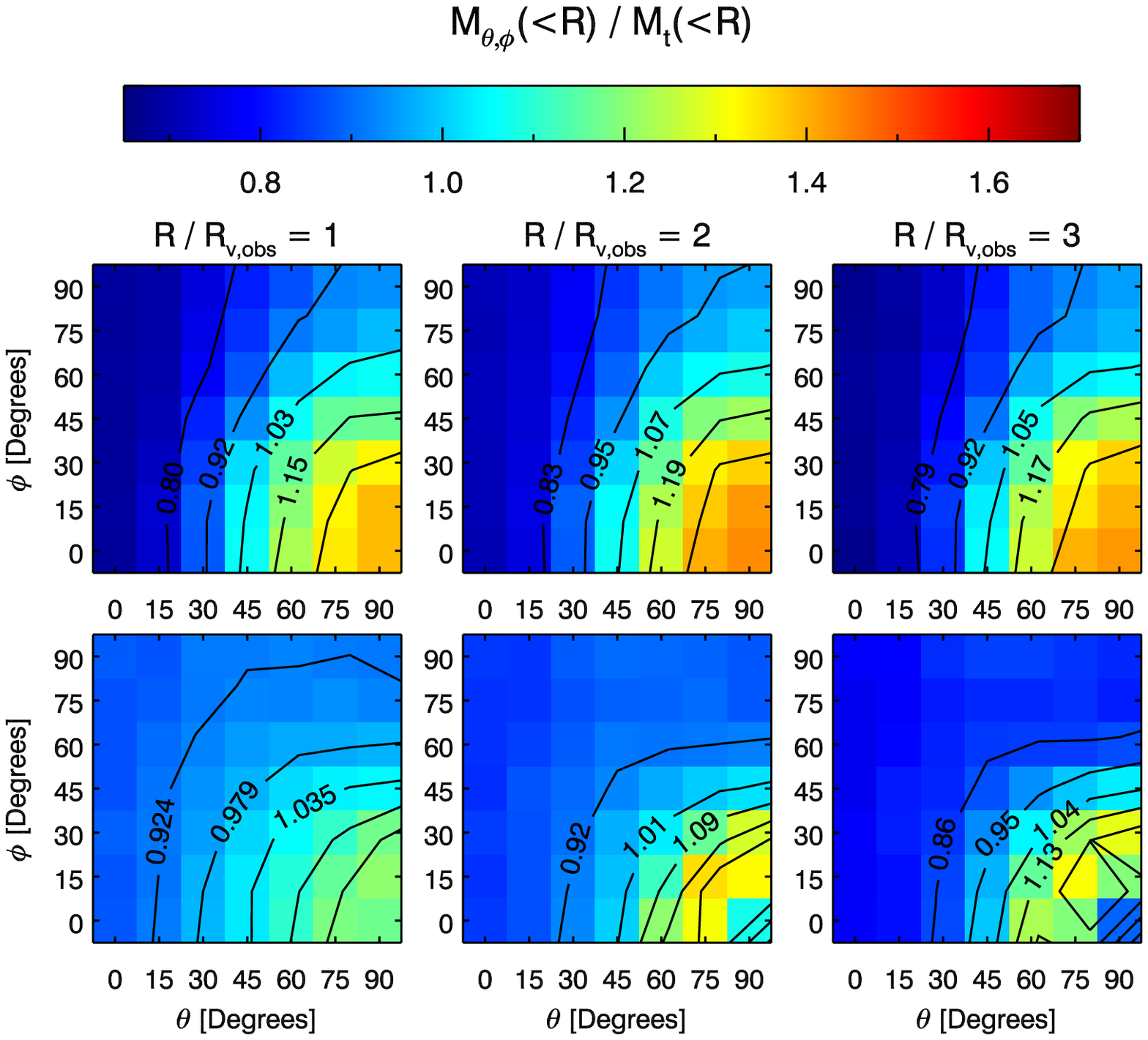}
		\caption{Effect of the cluster orientation on the caustic mass estimates for the low mass bin \textit{ellipsoidal} and \textit{filamentary stacks} evaluated at the observed virial radii $R_{v,obs}$. The panels show the mass estimates as a function of the orientation, relative to the true mass profile $M_t(<R)$. The three columns show results for three choices of radii. $\theta$ and $\phi$ indicate the line of sight in question, defined for each stack in Fig. \ref{fig:model}. For the top panels $(90^{\circ},0^{\circ})$ represents the sight line along the major axis, $(90^{\circ},90^{\circ})$ represents light line along minor axis and entire left side at $\theta=0^{\circ}$ represents the minor axis for any $\phi$. Note that the value of $R_{v,obs}$ differs with $(\theta,\phi)$, such that $M_{\theta,\phi}$ is evaluated at each of their respective $R_{v,obs}$, normalized to the true mass at the true $R_v$ The color for any $(\theta,\phi$) indicates the value of the mass estimate as indicated on the linear colorbar. The lines in each panel show equally-spaced isodensity contours.}
		\label{fig:obsrvir}
	\end{center}
\end{figure*}

The full angular dependency of the mass estimates from caustics is useful if cluster orientation is constrained. Tables \ref{tab:ellipsoid}, \ref{tab:filament}, \ref{tab:ellipsoidMB2}, and \ref{tab:filamentMB2} therefore show the values which are represented in Fig. \ref{fig:bias_both} and \ref{fig:bias_both2} for potential use in statistical calculations. The tables can also be used to estimate systematic errors in the caustic mass determination when cluster orientation is unknown or only partially constrained from observations. For the prolate and oblate substacks, Figs. \ref{fig:bias_obpro_low} and \ref{fig:bias_obpro_high} shows the angular dependency of the mass estimates from caustics on both the low and high mass bin. The two figures show how prolate and oblate clusters affect caustic mass estimates in the same manner, and are thus also virtually indistinguishable from the full stacks displayed in Fig. \ref{fig:bias_both} and \ref{fig:bias_both2} of the main article. This conclusion can also be drawn by comparing Table \ref{tab:results_oblate_prolate} that contains values for mass bias, scatter and major / minor axis values, to Table \ref{tab:results} of the main text. Finally, Fig. \ref{fig:obsrvir} shows the angular dependency of the caustic mass estimates, but using the iterative scheme for determining $R_v$ that is necessary when applying caustics to real data. This figure shows that caustics tend to over- and underestimate masses systematically slightly worse than when $R_v$ is considered a known parameter, as in our approach through most of the present work.

\begin{table}\begin{center}
    \caption{Caustic mass estimates of the low mass bin \textit{ellipsoidal stack} within 1$\,R_{v}$, 2$\,R_{v}$ and 3$\,R_{v}$ for varying line of sight, normalized by the true mass at these radii. Columns represent $\theta$ angles, rows represent $\phi$ angles, both are indicated in bold in the table and both are in degrees. These values are also represented as the top 3 panels of Fig. \ref{fig:bias_both}. All mass values are normalized by the true mass $M_t(<R)$.
\label{tab:ellipsoid}}
    \begin{tabular}{cccccccc}
$1\,R_{v}$& \textbf{0}     & \textbf{15}    & \textbf{30}    & \textbf{45} & \textbf{60}    & \textbf{75}    & \textbf{90}    \\   
\textbf{90} & 0.77 & 0.77 & 0.81 & 0.87 & 0.91 & 0.95 & 0.95 \\
\textbf{75} & 0.76 & 0.77 & 0.83 & 0.87 & 0.93 & 0.98 & 0.99 \\
\textbf{60} & 0.76 & 0.77 & 0.84 & 0.93 & 0.99 & 1.04 & 1.06 \\
\textbf{45} & 0.76 & 0.79 & 0.88 & 0.95 & 1.08 & 1.14 & 1.16 \\
\textbf{30} & 0.76 & 0.80 & 0.89 & 1.03 & 1.13 & 1.21 & 1.27 \\
\textbf{15} & 0.77 & 0.79 & 0.92 & 1.05 & 1.15 & 1.26 & 1.27 \\
\textbf{0} &  0.76 & 0.81 & 0.91 & 1.08 & 1.20 & 1.27 & 1.28 \\
\\
$2\,R_{v}$& \textbf{0}     & \textbf{15}    & \textbf{30}    & \textbf{45} & \textbf{60}    & \textbf{75}    & \textbf{90} \\   
\textbf{90} & 0.76 & 0.77 & 0.81 & 0.89 & 0.93 & 0.97 & 0.96 \\
\textbf{75} & 0.75 & 0.77 & 0.82 & 0.87 & 0.93 & 1.00 & 1.00 \\
\textbf{60} & 0.76 & 0.77 & 0.83 & 0.92 & 1.00 & 1.05 & 1.07 \\
\textbf{45} & 0.76 & 0.79 & 0.87 & 0.94 & 1.09 & 1.18 & 1.21 \\
\textbf{30} & 0.77 & 0.80 & 0.88 & 1.03 & 1.14 & 1.27 & 1.32 \\
\textbf{15} & 0.77 & 0.78 & 0.91 & 1.05 & 1.17 & 1.30 & 1.30 \\
\textbf{0} &  0.76 & 0.81 & 0.90 & 1.08 & 1.24 & 1.32 & 1.32 \\
\\
$3\,R_{v}$& \textbf{0}     & \textbf{15}    & \textbf{30}    & \textbf{45} & \textbf{60}    & \textbf{75}    & \textbf{90} \\    
\textbf{90} & 0.72 & 0.73 & 0.76 & 0.85 & 0.91 & 0.95 & 0.95 \\
\textbf{75} & 0.71 & 0.73 & 0.78 & 0.83 & 0.90 & 0.97 & 0.98 \\
\textbf{60} & 0.72 & 0.73 & 0.79 & 0.89 & 0.97 & 1.02 & 1.04 \\
\textbf{45} & 0.72 & 0.75 & 0.84 & 0.91 & 1.08 & 1.18 & 1.23 \\
\textbf{30} & 0.73 & 0.75 & 0.83 & 1.02 & 1.13 & 1.27 & 1.32 \\
\textbf{15} & 0.73 & 0.74 & 0.88 & 1.03 & 1.18 & 1.28 & 1.28 \\
\textbf{0} &  0.71 & 0.76 & 0.85 & 1.07 & 1.25 & 1.32 & 1.31 \\
    \end{tabular}
\end{center}\end{table}
\vspace*{0.3cm}

\begin{table}\begin{center}
    \caption{Caustic mass estimates of the low mass bin \textit{filamentary stack} within 1$\,R_{v}$, 2$\,R_{v}$ and 3$\,R_{v}$ for varying line of sight, normalized by the true mass at these radii. Columns represent $\theta$ angles, rows represent $\phi$ angles, both are indicated in bold in the table and both are in degrees. These values are also represented as the bottom 3 panels of Fig. \ref{fig:bias_both}. All mass values are normalized by the true mass $M_t(<R)$.
\label{tab:filament}}
    \begin{tabular}{cccccccc}
$1\,R_{v}$& \textbf{0}     & \textbf{15}    & \textbf{30}    & \textbf{45} & \textbf{60}    & \textbf{75}    & \textbf{90} \\ 
\textbf{90} & 0.89 & 0.90 & 0.94 & 0.95 & 0.96 & 0.96 & 0.94 \\
\textbf{75} & 0.91 & 0.91 & 0.94 & 0.95 & 0.95 & 0.96 & 0.92 \\
\textbf{60} & 0.89 & 0.92 & 0.96 & 0.97 & 0.97 & 1.02 & 0.98 \\
\textbf{45} & 0.90 & 0.93 & 0.94 & 1.01 & 1.03 & 1.04 & 1.05 \\
\textbf{30} & 0.91 & 0.91 & 0.98 & 1.01 & 1.06 & 1.10 & 1.15 \\
\textbf{15} & 0.89 & 0.92 & 0.98 & 1.05 & 1.07 & 1.15 & 1.17 \\
\textbf{0} &  0.91 & 0.92 & 0.98 & 1.04 & 1.08 & 1.15 & 1.12 \\
\\
$2\,R_{v}$& \textbf{0}     & \textbf{15}    & \textbf{30}    & \textbf{45} & \textbf{60}    & \textbf{75}    & \textbf{90} \\ 
\textbf{90} & 0.83 & 0.84 & 0.90 & 0.92 & 0.93 & 0.92 & 0.91 \\
\textbf{75} & 0.85 & 0.85 & 0.89 & 0.90 & 0.91 & 0.91 & 0.86 \\
\textbf{60} & 0.83 & 0.86 & 0.91 & 0.92 & 0.90 & 0.96 & 0.92 \\
\textbf{45} & 0.84 & 0.89 & 0.88 & 0.96 & 0.98 & 0.99 & 1.02 \\
\textbf{30} & 0.85 & 0.86 & 0.92 & 0.96 & 1.04 & 1.15 & 1.25 \\
\textbf{15} & 0.83 & 0.86 & 0.92 & 1.01 & 1.09 & 1.35 & 1.32 \\
\textbf{0} &  0.85 & 0.86 & 0.92 & 1.01 & 1.17 & 1.30 & 1.04 \\
\\
$3\,R_{v}$& \textbf{0}     & \textbf{15}    & \textbf{30}    & \textbf{45} & \textbf{60}    & \textbf{75}    & \textbf{90} \\ 
\textbf{90} & 0.76 & 0.79 & 0.85 & 0.87 & 0.89 & 0.88 & 0.86 \\
\textbf{75} & 0.79 & 0.79 & 0.83 & 0.85 & 0.85 & 0.85 & 0.79 \\
\textbf{60} & 0.77 & 0.80 & 0.85 & 0.86 & 0.84 & 0.90 & 0.87 \\
\textbf{45} & 0.77 & 0.83 & 0.81 & 0.90 & 0.95 & 0.99 & 1.03 \\
\textbf{30} & 0.78 & 0.80 & 0.86 & 0.92 & 1.04 & 1.21 & 1.28 \\
\textbf{15} & 0.77 & 0.79 & 0.86 & 1.00 & 1.13 & 1.34 & 1.21 \\
\textbf{0} &  0.79 & 0.80 & 0.85 & 1.00 & 1.20 & 1.20 & 0.86 \\
    \end{tabular}
\end{center}\end{table}
\vspace*{0.3cm}

\begin{table}\begin{center}
    \caption{Caustic mass estimates of the high mass bin \textit{ellipsoidal stack} within 1$\,R_{v}$, 2$\,R_{v}$ and 3$\,R_{v}$ for varying line of sight, normalized by the true mass at these radii. Columns represent $\theta$ angles, rows represent $\phi$ angles, both are indicated in bold in the table and both are in degrees. These values are also represented as the top 3 panels of Fig. \ref{fig:bias_both2}. All mass values are normalized by the true mass $M_t(<R)$.
\label{tab:ellipsoidMB2}}
    \begin{tabular}{cccccccc}
$1\,R_{v}$& \textbf{0}     & \textbf{15}    & \textbf{30}    & \textbf{45} & \textbf{60}    & \textbf{75}    & \textbf{90} \\ 
\textbf{90} & 0.75 & 0.78 & 0.81 & 0.88 & 0.92 & 0.98 & 0.97 \\
\textbf{75} & 0.74 & 0.79 & 0.81 & 0.89 & 0.94 & 0.99 & 1.04 \\
\textbf{60} & 0.76 & 0.78 & 0.84 & 0.92 & 1.02 & 1.11 & 1.14 \\
\textbf{45} & 0.74 & 0.80 & 0.87 & 0.99 & 1.09 & 1.21 & 1.26 \\
\textbf{30} & 0.75 & 0.79 & 0.91 & 1.08 & 1.21 & 1.39 & 1.43 \\
\textbf{15} & 0.75 & 0.82 & 0.92 & 1.12 & 1.29 & 1.47 & 1.48 \\
\textbf{0} &  0.76 & 0.81 & 0.93 & 1.11 & 1.35 & 1.46 & 1.49 \\
\\
$3\,R_{v}$& \textbf{0}     & \textbf{15}    & \textbf{30}    & \textbf{45} & \textbf{60}    & \textbf{75}    & \textbf{90} \\ 
\textbf{90} & 0.72 & 0.75 & 0.78 & 0.85 & 0.91 & 0.97 & 0.98 \\
\textbf{75} & 0.71 & 0.76 & 0.79 & 0.86 & 0.91 & 0.98 & 1.03 \\
\textbf{60} & 0.73 & 0.75 & 0.80 & 0.89 & 0.99 & 1.09 & 1.13 \\
\textbf{45} & 0.71 & 0.77 & 0.84 & 0.95 & 1.04 & 1.17 & 1.26 \\
\textbf{30} & 0.71 & 0.76 & 0.87 & 1.04 & 1.17 & 1.40 & 1.42 \\
\textbf{15} & 0.71 & 0.78 & 0.88 & 1.10 & 1.27 & 1.51 & 1.55 \\
\textbf{0} &  0.73 & 0.77 & 0.89 & 1.06 & 1.35 & 1.49 & 1.48 \\
\\
$3\,R_{v}$& \textbf{0}     & \textbf{15}    & \textbf{30}    & \textbf{45} & \textbf{60}    & \textbf{75}    & \textbf{90} \\ 
\textbf{90} & 0.66 & 0.70 & 0.73 & 0.79 & 0.87 & 0.93 & 0.95 \\
\textbf{75} & 0.65 & 0.70 & 0.73 & 0.81 & 0.87 & 0.94 & 0.99 \\
\textbf{60} & 0.68 & 0.70 & 0.75 & 0.84 & 0.95 & 1.04 & 1.07 \\
\textbf{45} & 0.66 & 0.72 & 0.78 & 0.90 & 0.99 & 1.13 & 1.20 \\
\textbf{30} & 0.66 & 0.70 & 0.81 & 0.99 & 1.14 & 1.41 & 1.42 \\
\textbf{15} & 0.66 & 0.73 & 0.82 & 1.05 & 1.24 & 1.53 & 1.55 \\
\textbf{0} &  0.68 & 0.72 & 0.83 & 1.02 & 1.34 & 1.53 & 1.50 \\
    \end{tabular}
\end{center}\end{table}
\vspace*{0.3cm}
\begin{table}\begin{center}
    \caption{Caustic mass estimatess of the high mass bin \textit{filamentary stack} within 1$\,R_{v}$, 2$\,R_{v}$ and 3$\,R_{v}$ for varying line of sight, normalized by the true mass at these radii. Columns represent $\theta$ angles, rows represent $\phi$ angles, both are indicated in bold in the table and both are in degrees. These values are also represented as the bottom 3 panels of Fig. \ref{fig:bias_both2}. All mass values are normalized by the true mass $M_t(<R)$.
\label{tab:filamentMB2}}
    \begin{tabular}{cccccccc}
$1\,R_{v}$& \textbf{0}     & \textbf{15}    & \textbf{30}    & \textbf{45} & \textbf{60}    & \textbf{75}    & \textbf{90} \\ 
\textbf{90} & 0.93 & 0.96 & 0.95 & 0.98 & 1.01 & 0.98 & 1.00 \\
\textbf{75} & 0.91 & 0.96 & 0.94 & 1.02 & 1.00 & 1.03 & 1.01 \\
\textbf{60} & 0.91 & 0.95 & 0.97 & 1.04 & 1.05 & 1.06 & 1.08 \\
\textbf{45} & 0.91 & 0.95 & 1.01 & 1.03 & 1.08 & 1.14 & 1.13 \\
\textbf{30} & 0.91 & 0.95 & 1.01 & 1.10 & 1.12 & 1.19 & 1.19 \\
\textbf{15} & 0.91 & 0.98 & 1.01 & 1.12 & 1.18 & 1.22 & 1.17 \\
\textbf{0} &  0.90 & 0.96 & 1.00 & 1.09 & 1.20 & 1.16 & 1.07 \\
\\
$3\,R_{v}$& \textbf{0}     & \textbf{15}    & \textbf{30}    & \textbf{45} & \textbf{60}    & \textbf{75}    & \textbf{90} \\ 
\textbf{90} & 0.89 & 0.90 & 0.89 & 0.92 & 0.97 & 0.94 & 0.97 \\
\textbf{75} & 0.86 & 0.91 & 0.89 & 0.97 & 0.95 & 1.00 & 0.98 \\
\textbf{60} & 0.86 & 0.91 & 0.92 & 1.00 & 0.99 & 0.99 & 1.03 \\
\textbf{45} & 0.85 & 0.90 & 0.96 & 0.96 & 1.01 & 1.08 & 1.08 \\
\textbf{30} & 0.86 & 0.90 & 0.95 & 1.03 & 1.06 & 1.15 & 1.18 \\
\textbf{15} & 0.87 & 0.94 & 0.95 & 1.06 & 1.16 & 1.27 & 1.16 \\
\textbf{0} &  0.85 & 0.91 & 0.95 & 1.02 & 1.16 & 1.17 & 0.96 \\
\\
$3\,R_{v}$& \textbf{0}     & \textbf{15}    & \textbf{30}    & \textbf{45} & \textbf{60}    & \textbf{75}    & \textbf{90} \\ 
\textbf{90} & 0.83 & 0.85 & 0.83 & 0.87 & 0.91 & 0.88 & 0.92 \\
\textbf{75} & 0.79 & 0.85 & 0.82 & 0.92 & 0.89 & 0.96 & 0.92 \\
\textbf{60} & 0.80 & 0.84 & 0.86 & 0.94 & 0.92 & 0.94 & 0.98 \\
\textbf{45} & 0.79 & 0.83 & 0.89 & 0.90 & 0.95 & 1.05 & 1.05 \\
\textbf{30} & 0.80 & 0.83 & 0.88 & 0.98 & 1.05 & 1.18 & 1.18 \\
\textbf{15} & 0.81 & 0.87 & 0.89 & 1.02 & 1.18 & 1.21 & 1.05 \\
\textbf{0} &  0.78 & 0.84 & 0.89 & 0.99 & 1.15 & 1.05 & 0.81 \\
    \end{tabular}
\end{center}\end{table}
\vspace*{0.3cm}

\begin{table*}\begin{center}
    \caption{Values of triaxiality and mass estimates for the oblate and prolate substacks of the \textit{ellipsoidal stacks}. The three 'Triaxiality' columns show the intermediate to major axis ratio $b/a$, the minor to major axis ratio $c/a$ and the triaxiality parameter $T$ for each of the stacks in questions for particles within $R_v$. The 'Mean mass' columns show the mean of 10,000 mass measurements obtained from 10,000 interpolations of each panel in Fig. \ref{fig:bias_both} and Fig. \ref{fig:bias_both2} for 10,000 directions randomly distributed on the sphere. The 'Scatter in masses' columns show the scatter defined as the standard deviation of the 10,000 mass measurements. The 'maj/min mass' columns show the ratio of the mass estimate obtained when observing along the major axis 
 to the mass estimate obtained when observing along the minor axis. All mass values are normalized by the true cluster mass $M_t(<R)$.
    \label{tab:results_oblate_prolate}}
    \begin{tabular}{ccccccccccccc}
    \toprule
    Stack name	& \multicolumn{3}{c}{Triaxiality} & \multicolumn{3}{c}{Mean mass} & \multicolumn{3}{c}{Scatter in masses} & \multicolumn{3}{c}{Maj/min mass} \\
     & $b/a$ & $c/a$ & $T$ & $1\,R_v$ & $2\,R_v$ & $3\,R_v$ & $1\,R_v$ & $2\,R_v$ & $3\,R_v$ & $1\,R_v$ & $2\,R_v$ & $3\,R_v$\\
    \hline \\   \vspace{2mm}	
    $M_v \in \left[ 1,2 \right]\times 10^{14}\,h^{-1}M_{\sun}$\\
    Prolate Stack 		& 	0.66	&	0.60	&	0.89	&	1.03&	1.03	&	1.01	&	0.14	&	0.13	& 0.14 &		1.65	&	1.68	&	1.76 \\ \vspace{2mm}
    Oblate Stack 		&	0.92	&	0.72	&	0.31	& 	1.02 & 1.07	& 1.06	&	0.15 &	0.17	&	0.15 & 1.80	&	1.80 & 1.70 \\ \vspace{2mm}
     $M_v \geq  2\times 10^{14}\,h^{-1}M_{\sun}$ & \\
    Prolate Stack 		&	0.66	&	0.61	&	0.89	&	1.09&	1.10	&	1.09	&	0.20	&	0.23	&	0.28	&	1.98	&	2.23	&	2.45\\
    Oblate Stack 		&	0.90	&	0.68	&	0.34	& 	1.07	&	1.07&	1.06	&	0.21&	0.24&	0.28	&	2.08	&	2.10&	2.16	\\
    \hline
    \end{tabular}
\end{center}\end{table*}


\end{document}